\documentclass[aps,prd,amsmath,superscriptaddress,twocolumn,floatfix,nofootinbib,preprintnumbers,10pt]{revtex4-2}

\usepackage{amssymb}\usepackage{txfonts}\usepackage{epsfig}\usepackage{bm}\usepackage{cancel}\usepackage{makecell}
\usepackage{xcolor}\usepackage{graphicx,graphics}\usepackage{multirow}\usepackage{float}\usepackage{ulem}
\usepackage{setspace}\usepackage{slashed}\usepackage{booktabs}\usepackage{xurl}\usepackage{hyperref}
\hypersetup{colorlinks=true, citecolor=blue, linkcolor=red, filecolor=black,urlcolor=blue, breaklinks=true}
\allowdisplaybreaks[4]

\newcommand{\eps}{\epsilon}
\newcommand{\Tr}{\operatorname{Tr}}
\newcommand{\hk}{\hat k}
\newcommand{\Lcal}{\mathcal L}
\newcommand{\bra}{\langle p|}
\newcommand{\ket}{|p\rangle}
\newcommand{\nbar}{\bar{n}}

\begin{document}

\title{Higher-twist hadronic tensor in deep inelastic scattering with massive quarks}

\author{Weihua Yang}\thanks{Corresponding author: yangwh@ytu.edu.cn}

\affiliation{College of Nuclear Equipment and Nuclear Engineering, Yantai University,\\ Yantai, Shandong 264005, China}

\author{Yongliang Yang}\thanks{Corresponding author: yangyl@qdu.edu.cn}
\affiliation{Centre for Theoretical and Computational Physics, College of Physics, Qingdao University, Qingdao, Shandong, 266071, China}
\author{Xiaoyu Wang}\thanks{Corresponding author: xiaoyuwang@zzu.edu.cn}
\affiliation{School of Physics, Zhengzhou University, Zhengzhou, Henan 450001, China}

\begin{abstract}

In this paper, we calculate the inclusive deep inelastic scattering (DIS) with finite quark masses up to twist-4 accuracy based on collinear expansion method. The collinear expansion method not only naturally leads to gauge-invariant correlation functions and parton distribution functions, but also yields a remarkably compact expression for the hadronic tensor. The resulting hadronic tensor can be organized according to the corresponding power behavior in $1/Q$. In particular, the explicit quark-mass dependence gives rise to additional power-suppressed contributions and the ratio $m/Q$  can be counted as a twist-3 contribution in the power expansion.  As an application, we further investigate the role of transversity in inclusive DIS with a transversely polarized target. We find that the transversity distribution appears explicitly in the quark-quark correlation function and gives rise to a contribution proportional to $m~h_{1T}/Q$. However, this contribution is exactly cancelled in the hadronic tensor by the one arising from the corresponding quark-gluon-quark correlation function. This cancellation shows that, within the one-photon-exchange approximation, the mass-induced transversity term cancels in the complete twist-3 hadronic tensor, and no net transversity contribution remains at twist-3 accuracy in inclusive DIS. We complement this analytical study with numerical estimates
of the $g_T$-related twist-3 asymmetry, comparing predictions
with and without the mass-dependent $x\rightarrow\chi$
rescaling at the same measured $x$.
The results show that these kinematic mass effects become
more important at lower $Q^2$ and larger $x$.

\end{abstract}

\maketitle

\section{Introduction} \label{sec:intro}

Deep-inelastic scattering (DIS) is one of the most powerful tools for probing the internal structure of nucleons and nuclei. Measurements of DIS have not only played a central role in establishing the partonic picture of the nucleon but also provided important information on both its momentum and spin structure \cite{Donnelly:1975ze,Bloom:1969kc,EuropeanMuon:1987isl,EuropeanMuon:1989yki}.  It will also remain an important process for exploring nucleon and nuclear structure at future facilities such as the Electron-Ion Collider (EIC) \cite{Accardi:2012qut,AbdulKhalek:2021gbh,Anderle:2021wcy}.
Thanks to the factorization theorem \cite{Collins:1989gx}, the DIS cross section can be factorized into a perturbatively calculable hard-scattering part and a nonperturbative soft part.
The nonperturbative information on the partonic structure of the nucleon is encoded in parton distribution functions (PDFs).  Different PDFs characterize different aspects of the nucleon, including the unpolarized, helicity, and transversity distribution functions, and so on. Consequently, measurable DIS cross sections can be systematically expressed as convolutions of perturbative hard-scattering parts with PDFs. Theoretical studies of DIS at the partonic level have been developed mainly within two complementary frameworks, the operator product expansion (OPE) \cite{Wilson:1969zs,Brandt:1970kg,Christ:1972ms} and the collinear expansion method \cite{Ellis:1982wd,Ellis:1982cd,Qiu:1990xxa,Qiu:1990xy,Qiu:1988dn}. The OPE organizes the DIS structure functions in terms of local operators and their Wilson coefficients, while the collinear expansion provides a diagrammatic framework in which the hard-scattering part is systematically expanded around the collinear partonic momentum. These methods provide complementary descriptions of the partonic structure and power-suppressed contributions in DIS, and form the theoretical basis for investigating higher-twist effects beyond the leading-twist approximation.

Compared with the OPE method, the collinear expansion offers a more transparent partonic picture and allows the hadronic tensor to be derived directly from gauge invariant correlation functions, even in the jet production semi-inclusive processes. 
However, in its conventional formulations and applications, the quark mass is usually neglected \cite{Liang:2006wp,Song:2010pf,Song:2013sja,Wei:2013csa,Wei:2014pma,Wei:2016far,Yang:2017sxz,Chen:2020ugq,Yang:2020ezt,Yang:2020qsk,Yang:2022mmh,Yang:2022knp,Yang:2022xwy}. This approximation is well justified for light quarks in many high-energy applications, but it becomes less adequate when mass-dependent power corrections are relevant. A finite quark mass can modify the hard part, the equations-of-motion relations, and the interplay between different correlation functions. It is therefore important to establish a systematic massive-quark formulation in which the quark mass is retained consistently throughout the collinear expansion.

In this paper, we develop such a framework for DIS with finite quark masses and calculate the hadronic tensor up to twist-4 accuracy. We find that the hard-scattering part of the hadronic tensor can be naturally separated into mass-dependent and mass-independent contributions, which are respectively convoluted with the chiral-odd and chiral-even components in the decomposition of the correlation functions. This separation provides a systematic way to identify the effects of the finite quark mass and the associated chiral-odd contributions in DIS. The resulting hadronic tensor can then be organized according to its power behavior in $1/Q$. In this expansion, the quark-mass dependence generates additional power-suppressed contributions and the ratio $m/Q$ is therefore counted as a twist-3 contribution as long as $m$ is on the order of $\Lambda_{QCD}$. 
As a particularly interesting application, we study the transversity \cite{Ralston:1979ys,Baldracchini:1980uq,Artru:1989zv} in inclusive DIS with a transversely polarized target. Transversity is a chiral-odd PDF and is well known to be absent from the usual leading-twist inclusive DIS because the electromagnetic interaction conserves chirality in the massless limit \cite{Jaffe:1991kp,Jaffe:1991ra,Jaffe:1997yz}. A nonzero quark mass, however, can provide a chirality-flipping mechanism and appears to allow a transversity contribution at twist-3 level \cite{Jaffe:1993xb,Boer:1996ay,Scopetta:1997qg,Jaffe:1997yz}. Indeed, within the massive-quark collinear expansion, we find that the quark-quark correlation function generates a term proportional to $m~h_{1T}/Q$. However, this term by itself does not satisfy gauge invariance. Once the corresponding quark-gluon-quark correlation function is included, its twist-3 contribution exactly cancels the transversity term from the quark-quark correlation function. This cancellation demonstrates that a finite quark mass does not generate a net transversity contribution to the inclusive DIS cross section at twist-3 accuracy. Nevertheless, to investigate the phenomenological relevance of twist-3 PDFs in DIS, we use a spectator model to estimate the contributions from both the $f_T$ term and the $g_T$ term. We find that $f_T$ term vanishes in the spectator model framework. Meanwhile the numerical estimates of the $g_T$ term indicate that the kinematic mass corrections become more
important at lower $Q^2$ and larger $x$. Furthermore, we notice that the predicted twist-3 asymmetry retains an appreciable magnitude despite its explicit $M/Q$ suppression, which allows to investigate twist-3 effects at EicC energy scale.


To be explicit, we organize this paper as follows. In Sec. \ref{sec:kinematics}, we first show the kinematics and conventions used to describe the inclusive DIS. In Sec. \ref{sec:hadron}, we present the complete expressions for the hadronic tensor including contributions up to two-gluon exchange in the parton model. In Sec. \ref{sec:collinear}, we present how to perform the collinear expansion with finite quark masses and obtain gauge-invariant correlation functions. In Sec. \ref{sec:applica}, we present the twist analysis and derive the cancellation of the transversity at twist-3 accuracy. We also present the numerical estimates for the $f_T$ and the $g_T$ terms. Finally, we present the summary in Sec. \ref{sec:sum}.

\section{Kinematics and conventions} \label{sec:kinematics}

We first provide a brief introduction of the kinematics and conventions used to describe the DIS process involving a longitudinally polarized electron scattering from a polarized proton target. We denote this process as 
\begin{align}
  e^-(l,\lambda_e) + N(p, S) \to e^-(l')+X, 
\end{align}
where $N$ denotes the proton with spin-1/2. Momenta for the relevant particles are shown in the parentheses.  $\lambda_e$ is the helicity of the incoming electron and $S$ is the spin of the proton.
The differential cross-section is written as 
\begin{align}
\frac{d\sigma}{dx dy} = \frac{\alpha^2 e_q^2}{2\pi s Q^2}\cdot L^{\mu\nu}(l,l')\cdot W_{\mu\nu}(p,q).   \label{f:diff}
\end{align}
Here $\alpha = e^2/4\pi$ is the fine structure constant, $e_q$ is the electric charge of the quark with flavor $q$. The leptonic tensor is given by
\begin{align}
L^{\mu\nu}(l,l') 
&=  2\left[ l^\mu l^{\prime\nu} + l^\nu l^{\prime\mu} - (l\cdot l^\prime)g^{\mu\nu}  \right] + 2i\lambda_e \varepsilon^{\mu\nu l l^\prime}. \label{f:leptont}
\end{align}
The operator definition of the hadronic tensor is given by
\begin{align}
W_{\mu\nu}(p,q) &= \sum_X \langle p|J_\mu^\dagger(0)|X\rangle\langle X|J_\nu(0)|p\rangle (2\pi)^4\delta^4(p+q-p_X)  \notag\\
&= \int d^4z  e^{iqz} \langle p|J_\mu^\dagger(0)\cdot J_\nu(z)|p\rangle. \label{f:hadronod}
\end{align}
The standard variables used in this paper are listed here,
\begin{align}
& s = (p+l)^2 = 2p\cdot l + M^2,  \qquad   M^2=p^2, \notag \\[4pt]
& y=\frac{p\cdot q}{p\cdot l},  \qquad  x=\frac{Q^2}{2p\cdot q}\notag \\[4pt]
& Q^2 = -q^2 = -(l-l')^2.
\end{align}
 
At the partonic level, it is convenient to use the light-cone variables to describe the DIS process. To be explicit, we introduce two light-cone unit vectors ($\bar n, n$),
\begin{align}
  & \bar{n}^\mu =\frac{1}{\sqrt{2}}(1, 0, 0, 1)_C = (1, 0, \vec 0_T)_L, \\
  & n^\mu =\frac{1}{\sqrt{2}}(1, 0, 0, -1)_C = (0, 1, \vec 0_T)_L.
\end{align}
Subscripts $C, L$ denote the Cartesian coordinate and the light-cone coordinate, respectively. These two vectors satisfy $\bar{n}^2=n^2=0$ and $\bar{n}\cdot n=1$.  With ($\bar n, n$), we can decompose the momentum $A$ as 
\begin{align}
  A^\mu=A^+ \bar{n}^\mu +A^- n^\mu +\vec{A}_T,
\end{align}
 where $A^\pm=A\cdot n (A\cdot \bar{n})$. The momentum square is  $A^2=2A^+A^- -\vec{A}_T^2$. 
 
 Under this convention, we can parametrize the momenta in the $\gamma^* N$ collinear frame as 
 \begin{align}
   & p^\mu =\left(p^+, \frac{M^2}{2p^+}, \vec{0}_T\right), \
   & q^\mu=\left(-\xi p^+, \frac{Q^2}{2\xi p^+},  \vec{0}_T\right),
 \end{align}
where $\xi$ is the Nachtmann variable,
\begin{align}
\xi = \frac{2x}{1+\sqrt{1+4x^2M^2/Q^2}}. \label{f:xi}
\end{align}
For the relevant quark,  we parametrize its momentum as 
\begin{align}
k^\mu = \hat x p^+\bar n^\mu + k_T^\mu + k^- n^\mu, 
\end{align}
where $\hat x = k^+/p^+$. Defining $\hat k^\mu = \hat x p^+\bar n^\mu$ will lead to
\begin{align}
k^\mu-\hat k^\mu = k_T^\mu + k^-n^\mu. 
\end{align}
We can rewrite the fraction $\hat{x}$ as
\begin{align}
\hat x = \xi\left(1+\frac{m^2}{Q^2}\right) \equiv \chi, \label{f:chib}
\end{align}
by using the on-shell condition of the scattered quark with mass $m$.
Equation \eqref{f:chib} show that the on-shell condition fixes the parton momentum fraction to $\hat x \to\chi$ when both the quark mass and the hadron mass are taken into account. Therefore, the momentum fraction entering the partonic hard scattering is modified from the usual Bjorken variable $x_B$ to the rescaled variable $\chi$, which incorporates both target-mass and quark-mass effects.  A systematic review of the mass corrections can be found in Ref. \cite{Ruiz:2023ozv}. 

\section{Hadronic tensor in the parton model}\label{sec:hadron}

\begin{figure}
  \centering
  \includegraphics[width=0.95\linewidth]{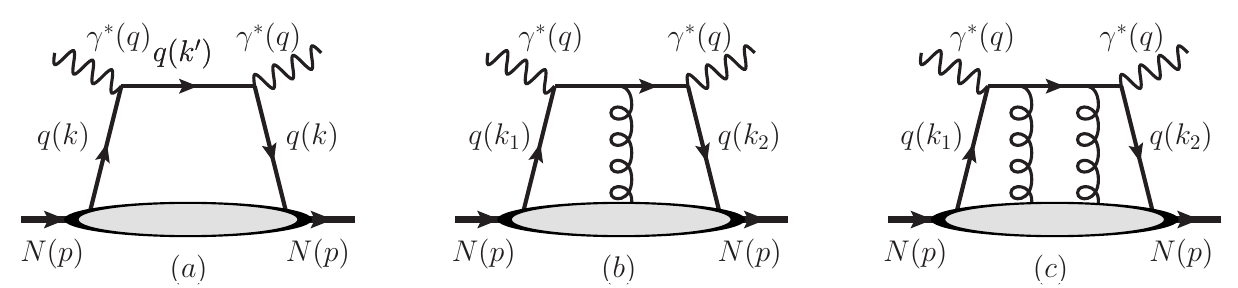}
  \caption{The first few diagrams of the Feynman diagram series with exchange of $j$ gluons, where $j=0, 1$, and $2$ for diagrams (a), (b), and (c), respectively. We emphasize that momenta are shown in parentheses and the quark symbol (q) should not be confused with the photon momentum (q). }\label{fig:dishad}
\end{figure}

Since the main focus of this work is to derive the collinear expansion with finite quark masses and explore its applications, we restrict ourselves here to the parton-model expression for the hadronic tensor and do not consider its representation in terms of structure functions \cite{Barone:2001sp,ParticleDataGroup:2026mpi}.
At the partonic level, DIS can be understood as the scattering of a virtual photon from an individual quark inside the nucleon under the one-photon exchange approximation, see Fig. \ref{fig:dishad}. The hadronic tensors can consequently be written as the convolution of  perturbatively calculable partonic hard-scattering coefficients with correlation functions describing the nonperturbative information on the partonic structure of the nucleon. To be explicit, we write the hadronic tensors as 
\begin{align}
  W_{\mu\nu}(p,q)=W_{\mu\nu}^{(0)}(p,q)+W_{\mu\nu}^{(1)}(p,q)+W_{\mu\nu}^{(2)}(p,q) +\cdots .
\end{align}
If we only consider the first few diagrams shown in Fig. \ref{fig:dishad}, the relevant hadronic tensors are given by
\begin{align}
 W_{\mu\nu}^{(0)}(p,q) &= \int \frac{d^4k}{(2\pi)^4} \Tr \left[\phi^{(0)}(p,k)  H_{\mu\nu}^{(0)}(q,k)\right], \label{f:w0} \\
 W_{\mu\nu}^{(1)} (p,q)&= \int \frac{d^4k_1}{(2\pi)^4}\frac{d^4k_2}{(2\pi)^4} \Tr \left[\phi_\rho^{(1)}(p,k_1,k_2)  H_{\mu\nu}^{(1)\rho}(q,k_1,k_2)\right],   \label{f:w1} \\
 W_{\mu\nu}^{(2)}(p,q) &= \int \frac{d^4k_1}{(2\pi)^4}\frac{d^4k}{(2\pi)^4}\frac{d^4k_2}{(2\pi)^4} \Tr \Big[\phi_{\rho\sigma}^{(2)}(p,k_1,k,k_2)\nonumber \\  
 &\qquad  \times H_{\mu\nu}^{(2)\rho\sigma}(q,k_1,k,k_2)\Big]. \label{f:w2} 
\end{align}
They respectively correspond to Fig. \ref{fig:dishad} (a), (b) and (c). Superscripts $0, 1$ and $2$ denote the number of gluons, $j=0, 1$, and $2$.  The hard-scattering tensors (from now on we call the hard-scattering coefficients as tensors) including quark masses are given by
\begin{widetext}
\begin{align}
H_{\mu\nu}^{(0)}(q,k) &= \gamma_\mu(\slashed k+\slashed q+m)\gamma_\nu (2\pi) \delta \left[(k+q)^2-m^2\right],  \label{f:hmunu0} \\[2pt]
H_{\mu\nu}^{(1)\rho} (q,k_1,k_2)&= H_{\mu\nu}^{(1)L,\rho} + H_{\mu\nu}^{(1)R,\rho} \notag\\[2pt]
&= \gamma_\mu\frac{(\slashed k_2+\slashed q+m)}{(k_2+q)^2-m^2-i\eps}\gamma^\rho(\slashed k_1+\slashed q+m)\gamma_\nu (2\pi)\delta \left[(k_1+q)^2-m^2\right] \notag\\[2pt]
&+\gamma_\mu(\slashed k_2+\slashed q+m)\gamma^\rho\frac{(\slashed k_1+\slashed q+m)}{(k_1+q)^2-m^2+i\eps}\gamma_\nu (2\pi)\delta \left[(k_2+q)^2-m^2\right],  \label{f:hmunu1} \\[2pt]
H_{\mu\nu}^{(2)\rho\sigma} (p,k_1,k,k_2)&= H_{\mu\nu}^{(2)L,\rho\sigma} + H_{\mu\nu}^{(2)M,\rho\sigma} + H_{\mu\nu}^{(2)R,\rho\sigma} \notag\\[2pt]
&= \gamma_\mu\frac{(\slashed k_2+\slashed q+m)}{(k_2+q)^2-m^2-i\eps}\gamma^\rho\frac{(\slashed k+\slashed q+m)}{(k+q)^2-m^2-i\eps}\gamma^\sigma(\slashed k_1+\slashed q+m)\gamma_\nu (2\pi)\delta \left[(k_1+q)^2-m^2\right] \notag\\[2pt]
&+\gamma_\mu\frac{(\slashed k_2+\slashed q+m)}{(k_2+q)^2-m^2-i\eps}\gamma^\rho(\slashed k+\slashed q+m)\gamma^\sigma\frac{(\slashed k_1+\slashed q+m)}{(k_1+q)^2-m^2+i\eps}\gamma_\nu (2\pi)\delta \left[(k+q)^2-m^2\right] \notag\\[2pt]
&+\gamma_\mu(\slashed k_2+\slashed q+m)\gamma^\rho\frac{(\slashed k+\slashed q+m)}{(k+q)^2-m^2+i\eps}\gamma^\sigma\frac{(\slashed k_1+\slashed q+m)}{(k_1+q)^2-m^2+i\eps}\gamma_\nu (2\pi)\delta \left[(k_2+q)^2-m^2\right]. \label{f:hmunu2}
\end{align}
\end{widetext}
We note that $ H_{\mu\nu}^{(1)L,\rho}$ and $H_{\mu\nu}^{(1)R,\rho}$ respectively denote the left-cut and right-cut hadronic tensors for $j=1$ case. Left-(right-)cut implies the cut line is  on the left (right) side of the gluon in  Fig. \ref{fig:dishad} (b).  The similar conventions also apply to $H_{\mu\nu}^{(2)L,\rho\sigma}, H_{\mu\nu}^{(2)M,\rho\sigma}$ and $H_{\mu\nu}^{(2)R,\rho\sigma}$ in Fig. \ref{fig:dishad} (c).  The corresponding correlation functions are defined as
\begin{align}
\phi^{(0)}(p,k) &= \int d^4z  e^{ikz} \bra\bar\psi(0)\psi(z)\ket,   \label{f:phi0} \\
\phi_\rho^{(1)}(p,k_1,k_2) &= \int d^4z_1 d^4z_2  e^{ik_1(z_1-z_2)}e^{ik_2z_2}\notag \\
&\times \bra\bar\psi(0)A_\rho(z_2)\psi(z_1)\ket,  \label{f:phi1} \\
\phi_{\rho\sigma}^{(2)}(p,k_1,k,k_2) &= \int d^4z_1 d^4z d^4z_2\; e^{ik_1(z_1-z)}e^{ik(z-z_2)}e^{ik_2z_2}\notag \\
&\times \bra\bar\psi(0)A_\rho(z)A_\sigma(z_2)\psi(z_1)\ket,  \label{f:phi2}
\end{align}
where the strong coupling constant $g$ is omitted in these correlation functions for brevity. Since the gauge links are not explicitly included, these correlation functions are not manifestly gauge invariant. In the following section, we demonstrate how gauge-invariant correlation functions naturally emerge within the collinear expansion method.

\section{Collinear expansion with finite quark mass} \label{sec:collinear}

\subsection{Taylor expansion of the hard-scattering tensors }

The collinear expansion method, also known as the EFP method \cite{Ellis:1982wd,Ellis:1982cd}, provides a systematic and diagrammatic framework for calculating the DIS hadronic tensor beyond leading twist.  In this method, the hard-scattering tensor is expanded around the collinear momentum, $\hat{k}=\chi p^+$. The transverse terms are transferred to the soft part through integrations by parts. In this way, derivatives of the quark fields are combined with the gluon fields into covariant derivatives, while the longitudinal gluon contributions are resummed into gauge links. The resulting correlation functions are therefore gauge invariant and can be classified clearly. The hadronic tensor is consequently expressed as a systematic sum of hard-scattering tensor convoluted with gauge-invariant correlation functions at successive twist orders. This procedure provides a transparent connection between the momentum expansion of the hard-scattering tensors and the emergence of gauge-invariant correlation functions. In the following, we provide a detailed derivation of the collinear expansion with finite quark mass.


We first make a Taylor expansion of all the hard-scattering tensors around $\hat{k}=\chi p^+$. 
 \begin{align}
  H_{\mu\nu}^{(0)}(k) &= H_{\mu\nu}^{(0)}(\hk) + \frac{\partial H_{\mu\nu}^{(0)}(\hk)}{\partial \hk_\rho}(k-\hk)_\rho \notag \\
& +\frac12\frac{\partial^2H_{\mu\nu}^{(0)}(\hk)}{\partial\hk_\rho\partial\hk_\sigma}(k-\hk)_\rho(k-\hk)_\sigma+\cdots,   \label{f:hmunu0ex}\\
H_{\mu\nu}^{(1)i,\rho}(k_1,k_2) &= H_{\mu\nu}^{(1)i,\rho}(\hat k_1,\hat k_2) + \frac{\partial H_{\mu\nu}^{(1)i,\rho}(\hat k_1,\hat k_2)}{\partial \hat k_{1}{}_\rho}(k_1-\hat k_1)_\rho\notag \\
&  +\frac{\partial H_{\mu\nu}^{(1)i,\rho}(\hat k_1,\hat k_2)}{\partial\hat k_{2}{}_\rho}(k_2-\hat k_2)_\rho + \cdots, \label{f:hmunu1ex} \\[4pt]
H_{\mu\nu}^{(2)i,\rho\sigma}(k_1,k,k_2)& = H_{\mu\nu}^{(2)i,\rho\sigma}(\hat k_1,\hat k,\hat k_2)+\cdots. \label{f:hmunu2ex}
\end{align}
In Eqs. \eqref{f:hmunu1ex} and \eqref{f:hmunu2ex}, the superscript $i=L,R$ and $i=L,M,R$, respectively. We use the Ward identity to relate the collinear expansion of the lower-$g$ hard-scattering tensors to the higher-$g$ hard-scattering tensors, and obtain the following relations,
\begin{align}
\frac{\partial H_{\mu\nu}^{(0)}(k)}{\partial k_\rho} &= -\left[H_{\mu\nu}^{(1)L,\rho}(k,k)+H_{\mu\nu}^{(1)R,\rho}(k,k)\right], \tag{3-22}\\[4pt]
\frac{\partial H_{\mu\nu}^{(1)L,\rho}(k_1,k_2)}{\partial k_{1\sigma}} &= -\left[H_{\mu\nu}^{(2)L,\rho\sigma}+H_{\mu\nu}^{(2)M,\rho\sigma}\right](k_1,k_1,k_2), \tag{3-23}\\[4pt]
\frac{\partial H_{\mu\nu}^{(1)R,\rho}(k_1,k_2)}{\partial k_{1\sigma}} &= -H_{\mu\nu}^{(2)R,\rho\sigma}(k_1,k_1,k_2), \tag{3-24}\\[4pt]
\frac{\partial H_{\mu\nu}^{(1)L,\rho}(k_1,k_2)}{\partial k_{2\sigma}} &= -H_{\mu\nu}^{(2)L,\rho\sigma}(k_1,k_2,k_2), \tag{3-25}\\[4pt]
\frac{\partial H_{\mu\nu}^{(1)R,\rho}(k_1,k_2)}{\partial k_{2\sigma}} &= -\left[H_{\mu\nu}^{(2)M,\rho\sigma}+H_{\mu\nu}^{(2)R,\rho\sigma}\right](k_1,k_2,k_2), \tag{3-26}\\[4pt]
\frac{\partial^2 H_{\mu\nu}^{(0)}(k)}{\partial k_{1\rho}\partial k_{2\sigma}} &= \sum_{i=L,M,R}\Big[H_{\mu\nu}^{(2)i,\rho\sigma}+H_{\mu\nu}^{(2)i,\sigma\rho}\Big](k,k,k). \tag{3-27}
\end{align}
Here we have used
\begin{align}
 & \frac{\partial}{\partial k_\rho}\left(\frac{\slashed k+m}{k^2-m^2\pm i\eps}\right) = -\frac{\slashed k+m}{k^2-m^2\pm i\eps}\gamma^\rho\frac{\slashed k+m}{k^2-m^2\pm i\eps}, \\[6pt]
 & \frac{\partial}{\partial k_\rho}\Bigg\{(\slashed k+\slashed q+m) \delta\Big[(k+q)^2-m^2\Big]\Bigg\} \notag \\
 &= -\frac{\slashed k+\slashed q+m}{(k+q)^2-m^2}\gamma^\rho(\slashed k+\slashed q+m) \delta\Big[(k+q)^2-m^2\Big] \notag\\
&\quad-(\slashed k+\slashed q+m)\gamma^\rho\frac{\slashed k+\slashed q+m}{(k+q)^2-m^2} \delta\Big[(k+q)^2-m^2\Big]. \label{f:partial}
\end{align}

Hence, we can rewrite the Taylor expansion of all the hard-scattering tensors,
  \begin{align}
H_{\mu\nu}^{(0)}(k) &= H_{\mu\nu}^{(0)}(\hk) - \left[H_{\mu\nu}^{(1)L,\rho}+H_{\mu\nu}^{(1)R,\rho}\right](\hk,\hk)\omega_\rho^{\rho'}k_{\rho'} \notag\\[4pt]
&+\Big[H_{\mu\nu}^{(2)L,\rho\sigma}+H_{\mu\nu}^{(2)M,\rho\sigma}+H_{\mu\nu}^{(2)R,\rho\sigma}\Big](\hk,\hk,\hk)\notag \\[4pt]
&\times \omega_\rho^{\rho'} \omega_\sigma^{\sigma'}k_{\rho'}k_{\sigma'}+\cdots,  \label{f:h0ex}\\[4pt]
H_{\mu\nu}^{(1)L,\rho}(k_1,k_2) &= H_{\mu\nu}^{(1)L,\rho}(\hat k_1,\hat k_2) -H_{\mu\nu}^{(2)L,\rho\sigma}(\hat k_1,\hat k_2,\hat k_2)\omega_\sigma^{\sigma'}k_{2\sigma'}  \notag \\[4pt]
&- \left[H_{\mu\nu}^{(2)L,\rho\sigma}+H_{\mu\nu}^{(2)M,\rho\sigma}\right](\hat k_1,\hat k_1,\hat k_2)\omega_\sigma^{\sigma'}k_{1\sigma'} \notag\\[4pt]
&+\cdots, \\[4pt]
H_{\mu\nu}^{(1)R,\rho}(k_1,k_2) &= H_{\mu\nu}^{(1)R,\rho}(\hat k_1,\hat k_2) - H_{\mu\nu}^{(2)R,\rho\sigma}(\hat k_1,\hat k_1,\hat k_2)\omega_\sigma^{\sigma'}k_{1\sigma'} \notag \\[4pt]
& -\left[H_{\mu\nu}^{(2)M,\rho\sigma}+H_{\mu\nu}^{(2)R,\rho\sigma}\right](\hat k_1,\hat k_2,\hat k_2)\omega_\sigma^{\sigma'}k_{2\sigma'} \notag\\[4pt]
&+\cdots,  \\[4pt]
H_{\mu\nu}^{(2)i,\rho\sigma}(k_1,k, &k_2) = H_{\mu\nu}^{(2)i,\rho\sigma}(\hat k_1,\hat k,\hat k_2)+\cdots. \label{f:h2ex}
\end{align}
Here $\omega_\rho^{\rho'}$ is the projection operator which is defined as, $\omega_\rho^{\rho'}=g^{\rho'}_\rho-\bar{n}_\rho n^{\rho'}$ \cite{Liang:2006wp}. Therefore we have $\omega_\rho^{\rho'}k_{\rho'}=(k-\hat{k})_\rho$.

\subsection{Hadronic tensors after collinear expansion}

  Next, we substitute Eqs. \eqref{f:h0ex}-\eqref{f:h2ex} into hadronic tensors shown in Eqs. \eqref{f:w0}-\eqref{f:w2} and obtain the new tensors. For $W_{\mu\nu}^{(0)}$, we have  
  \begin{align}
W_{\mu\nu}^{(0)} &= \int_{k_1}\Tr\big[\phi^{(0)}(k_1)H_{\mu\nu}^{(0)}(\hat k_1)\big] \notag\\
&-\sum_{i=1}^{2}\int_{k_1,k_2}\Tr\big[\phi_\rho^{(0)i}(k_1,k_2)H_{\mu\nu}^{(1)i,\rho}(\hat k_1,\hat k_2)\big]\omega_\rho^{\rho'} \notag\\
&+\sum_{i=1}^{3}\int_{k_1,k,k_2}\Tr\big[\phi_{\rho'\sigma'}^{(0)}(k_1,k,k_2)H_{\mu\nu}^{(2)i,\rho\sigma}(\hat k_1,\hat k,\hat k_2)\big]\omega_\rho^{\rho'}\omega_\sigma^{\sigma'},
\end{align}
where $\int_{k_1}$ is the shorthand notation of $\int\frac{d^4k_1}{(2\pi)^4}$, i.e. $\int_{k_1}\equiv\int\frac{d^4k_1}{(2\pi)^4}$, and $\sum_{i=1}^{2}$ denotes the summation over \(L\) and \(R\), while $ \sum_{i=1}^{3} $
denotes the summation over \(L\), \(M\), and \(R\). The redefined correlation functions are 
\begin{align}
\phi_{\rho}^{(0)}(k_1,k_2) &= \int d^4z_1 d^4z_2  e^{ik_1(z_1-z_2)}e^{ik_2z_2}\notag \\
&\times \bra\bar\psi(0) i\partial_{\rho}\psi(z_1)\ket,  \\
\phi_{\rho\sigma}^{(0)}(k_1,k,k_2) &= \int d^4z_1 d^4z d^4z_2  e^{ik_1(z_1-z)}e^{ik_2z_2}e^{ik(z-z_2)}\notag \\
&\times  \bra\bar\psi(0) i\partial_{\rho} i\partial_{\sigma}\psi(z_1)\ket. 
\end{align}
For $W_{\mu\nu}^{(1)}$, we have  
\begin{align}
W_{\mu\nu}^{(1)} &=\int_{k_1,k_2}\Bigg\{
\sum_{i=1}^{2}\Tr\Big[\frac{n\cdot\phi^{(1)}(k_1,k_2)}{p\cdot n}p_\rho H_{\mu\nu}^{(1)i,\rho}(\hat k_1,\hat k_2)\Big] \notag\\
&+\sum_{i=1}^{2}\Tr\big[\phi_{\rho'}^{(1)}(k_1,k_2)H_{\mu\nu}^{(1)i,\rho}(\hat k_1,\hat k_2)\big]\omega_\rho^{\rho'} \notag\\
&-\sum_{i=1}^{3}\Tr\big[\phi_{\rho'}^{(1)}(k_1,k_2)H_{\mu\nu}^{(2)i,\rho\sigma}(\hat k_1,\hat k_1,\hat k_2)\big]\omega_\sigma^{\sigma'}k_{1\sigma'}\omega_\rho^{\rho'} \notag\\
&-\sum_{i=1}^{3}\Tr\big[\phi_{\rho'}^{(1)}(k_1,k_2)H_{\mu\nu}^{(2)i,\sigma\rho}(\hat k_1,\hat k_2,\hat k_2)\big]\omega_\sigma^{\sigma'}k_{2\sigma'}\omega_\rho^{\rho'} \notag\\
&-\sum_{i=1}^{3}\Tr\Big[\frac{n\cdot\phi^{(1)}(k_1,k_2)}{p\cdot n}p_\rho H_{\mu\nu}^{(2)i,\rho\sigma}(\hat k_1,\hat k_1,\hat k_2)\Big]\omega_\sigma^{\sigma'}k_{1\sigma'} \notag\\
&-\sum_{i=1}^{3}\Tr\Big[\frac{n\cdot\phi^{(1)}(k_1,k_2)}{p\cdot n}p_\rho H_{\mu\nu}^{(2)i,\sigma\rho}(\hat k_1,\hat k_2,\hat k_2)\Big]\omega_\sigma^{\sigma'}k_{2\sigma'}\Bigg\}.   
\end{align}
In obtaining this formula, we have divided the correlation function into the longitudinal and transverse parts, 
\begin{align}
\phi_\rho^{(1)}(k_1,k_2) = \omega_\rho^{\rho'}\phi_{\rho'}^{(1)}(k_1,k_2) + \frac{n\cdot\phi^{(1)}(k_1,k_2)}{n\cdot p}p_\rho.  \label{f:phi1dec}
\end{align} 
By absorbing the quark momentum into the correlation function, we obtain
\begin{align}
W_{\mu\nu}^{(1)} &=\int_{k_1,k_2}\Bigg\{
\sum_{i=1}^{2}\Tr\Big[\frac{n\cdot\phi^{(1)}(k_1,k_2)}{p\cdot n}p_\rho H_{\mu\nu}^{(1)i,\rho}(\hat k_1,\hat k_2)\Big] \notag\\
&+\sum_{i=1}^{2}\Tr\big[\phi_{\rho'}^{(1)}(k_1,k_2)H_{\mu\nu}^{(1)i,\rho}(\hat k_1,\hat k_2)\big]\omega_\rho^{\rho'} \notag\\
&-\sum_{i=1}^{3}\Tr\big[\phi_{1\rho'\sigma'}^{(1)}(k_1,k_2)H_{\mu\nu}^{(2)i,\rho\sigma}(\hat k_1,\hat k_1,\hat k_2)\big]\omega_\sigma^{\sigma'}\omega_\rho^{\rho'} \notag\\
&-\sum_{i=1}^{3}\Tr\big[\phi_{2\rho'\sigma'}^{(1)}(k_1,k_2)H_{\mu\nu}^{(2)i,\sigma\rho}(\hat k_1,\hat k_2,\hat k_2)\big]\omega_\sigma^{\sigma'}\omega_\rho^{\rho'} \notag\\
&-\sum_{i=1}^{3}\Tr\Big[\frac{n\cdot\phi^{(1)}_{1\sigma'}(k_1,k_2)}{p\cdot n}p_\rho H_{\mu\nu}^{(2)i,\rho\sigma}(\hat k_1,\hat k_1,\hat k_2)\Big]\omega_\sigma^{\sigma'}\notag\\
&-\sum_{i=1}^{3}\Tr\Big[\frac{n\cdot\phi^{(1)}_{2\sigma'}(k_1,k_2)}{p\cdot n}p_\rho H_{\mu\nu}^{(2)i,\sigma\rho}(\hat k_1,\hat k_2,\hat k_2)\Big]\omega_\sigma^{\sigma'}\Bigg\},\label{f:w1ex}
\end{align}
where the relevant correlation functions  are defined as 
\begin{align}
\frac{n\cdot\phi^{(1)}(k_1,k_2)}{p\cdot n} &= \int d^4z_1 d^4z_2 e^{ik_1(z_1-z_2)}e^{ik_2z_2}\nonumber \\
&\times  \bra\bar\psi(0) \frac{A\cdot n}{p\cdot n}(z_2) \psi(z_1)\ket,  \\
\frac{n\cdot\phi_{1\sigma'}^{(1)}(k_1,k,k_2)}{p\cdot n} &= \int d^4z_1 d^4z d^4z_2 e^{ik_1(z_1-z)}e^{ik_2z_2}e^{ik(z_-z_2)}\nonumber \\
&\times  \bra\bar\psi(0) \frac{n\cdot A(z_2)}{p\cdot n} i\partial_{\sigma'}\psi(z_1)\ket, \\
\frac{n\cdot\phi_{2\sigma'}^{(1)}(k_1,k,k_2)}{p\cdot n} &= \int d^4z_1 d^4z d^4z_2\; e^{ik_1(z_1-z)}e^{ik(z-z_2)}e^{ik_2z_2}\notag \\
\times \bra\bar\psi(0)&\Big[i\partial_{\sigma'}(z_2)+i\partial_{\sigma'}(z_1)\Big]\frac{n\cdot A(z_2)}{p\cdot n}\psi(z_1)\ket, \\
\phi_{1\rho'\sigma'}^{(1)}(k_1,k,k_2) &=\int d^4z_1 d^4z d^4z_2\; e^{ik_1(z_1-z)}e^{ik(z-z_2)}e^{ik_2z_2}\notag \\[2pt]
&\times \bra\bar\psi(0)A_{\rho'}(z_2) i\partial_{\sigma'}\psi(z_1)\ket, \\[4pt]
\phi_{2\rho'\sigma'}^{(1)}(k_1,k,k_2) &= \int d^4z_1 d^4z d^4z_2\; e^{ik_1(z_1-z)}e^{ik(z-z_2)}e^{ik_2z_2}\notag \\
\times \bra \bar\psi&(0)\Big[i\partial_{\sigma'}(z_2)+i\partial_{\sigma'}(z_1)\Big]A_{\rho'}(z_2)\psi(z_1)\ket.
\end{align}

Following the decomposition in Eq. \eqref{f:phi1dec}, we can decompose $\phi_{\rho\sigma}^{(2)}(k_1,k,k_2)$ as well, and rewrite the  hadronic tensor $W_{\mu\nu}^{(2)}$ in the following form,
\begin{align}
W_{\mu\nu}^{(2)} &=\int_{k_1,k,k_2}\sum_{i=1}^{3}\Bigg\{
\Tr\big[\phi_{\rho'\sigma'}^{(2)}(k_1,k,k_2)H_{\mu\nu}^{(2)i,\rho\sigma}(\hat k_1,\hat k,\hat k_2)\big]\omega_\rho^{\rho'}\omega_\sigma^{\sigma'} \notag\\[4pt]
&+\Tr\Big[\Big(\omega_\rho^{\rho'}\frac{n^{\sigma'}\phi_{\rho'\sigma'}^{(2)}(k_1,k,k_2)}{p\cdot n}p_\sigma + \omega_\sigma^{\sigma'}\frac{n^{\rho'}\phi_{\rho'\sigma'}^{(2)}(k_1,k,k_2)}{p\cdot n}p_\rho\Big)\nonumber \\[4pt]
&\qquad \times H_{\mu\nu}^{(2)i,\rho\sigma}(\hat k_1,\hat k,\hat k_2)\Big] \notag\\[4pt]
&+\Tr\Big[\frac{n^{\rho'}n^{\sigma'}\phi_{\rho'\sigma'}^{(2)}(k_1,k,k_2)}{(p\cdot n)^2}p_\rho p_\sigma H_{\mu\nu}^{(2)i,\rho\sigma}(\hat k_1,\hat k,\hat k_2)\Big]\Bigg\}. 
\end{align}
Here $\phi_{\rho\sigma}^{(2)}(k_1,k,k_2)$ has been shown in Eq. \eqref{f:phi2}.

Next, we use the on-shell condition to simplify the hard-scattering tensors and obtain the corresponding simplified hadronic tensor. The simplified hard-scattering tensors are given by
  \begin{align}
&   H_{\mu\nu}^{(0)}(\hk)  =\gamma_\mu(\slashed{\hk}+\slashed q+m)\gamma_\nu \frac{1}{2p\cdot q}(2\pi)\delta(x-\chi), \\[4pt] 
&   H_{\mu\nu}^{(1)L,\rho}(\hk_1,\hk_2) =\gamma_\mu\frac{\slashed{\hk}_2+\slashed q+m}{x_2-\chi-i\eps}\gamma^\rho(\slashed{\hk}_1+\slashed q+m)\gamma_\nu  \nonumber \\ 
& \qquad \qquad  \qquad \times\frac{2\pi}{(2p\cdot q)^2}\delta(x_1-\chi),\\[4pt]
&    H_{\mu\nu}^{(1)R,\rho}(\hk_1,\hk_2) = \gamma_\mu(\slashed{\hk}_2+\slashed q+m)\gamma^\rho \frac{\slashed{\hk}_1+\slashed q+m}{x_1-\chi+i\eps}\gamma_\nu  \nonumber \\ 
& \qquad \qquad  \qquad \times  \frac{2\pi}{(2p\cdot q)^2}\delta(x_2-\chi),\\[4pt]
&   H_{\mu\nu}^{(2)L,\rho\sigma}(\hk_1,\hk,\hk_2) = \gamma_\mu\frac{\slashed{\hk}_2+\slashed q+m}{x_2-\chi-i\eps}\gamma^\rho\frac{\slashed{\hk}+\slashed q+m}{x-\chi-i\eps}\gamma^\sigma\nonumber \\ 
& \qquad \qquad  \qquad \times (\slashed{\hk}_1+\slashed q+m)\gamma_\nu \frac{2\pi}{(2p\cdot q)^3}\delta(x_1-\chi),\\[4pt]
&   H_{\mu\nu}^{(2)M,\rho\sigma}(\hk_1,\hk,\hk_2) = \gamma_\mu\frac{\slashed{\hk}_2+\slashed q+m}{x_2-\chi-i\eps}\gamma^\rho(\slashed{\hk}+\slashed q+m)\gamma^\sigma\nonumber \\ 
& \qquad \qquad  \qquad \times \frac{\slashed{\hk}_1+\slashed q+m}{x_1-\chi+i\eps}\gamma_\nu \frac{2\pi}{(2p\cdot q)^3}\delta(x-\chi),\\[4pt]
&   H_{\mu\nu}^{(2)R,\rho\sigma}(\hk_1,\hk,\hk_2) = \gamma_\mu(\slashed{\hk}_2+\slashed q+m)\gamma^\rho\frac{\slashed{\hk}+\slashed q+m}{x-\chi+i\eps}\gamma^\sigma \nonumber \\ 
& \qquad \qquad  \qquad \times\frac{\slashed{\hk}_1+\slashed q+m}{x_1-\chi+i\eps}\gamma_\nu \frac{2\pi}{(2p\cdot q)^3}\delta(x_2-\chi). 
\end{align}
We can also absorb the momentum $p$ into the hard-scattering tensor and use this equation,
  \begin{align}
 &\frac{\slashed q+\slashed{\hk}_2+m}{x_2-\chi-i\eps}\slashed p \frac{\slashed q+\slashed{\hk}_1+m}{x_1-\chi-i\eps} \notag \\
=& 2p\cdot q\Bigg[\frac{1}{x_2-x_1-i\eps}\cdot\frac{\slashed q+\slashed{\hk}_1+m}{x_1-\chi-i\eps} \notag\\
&\qquad \quad+\frac{1}{x_1-x_2+i\eps}\cdot\frac{\slashed q+\slashed{\hk}_2+m}{x_2-\chi-i\eps}\Bigg]. 
\end{align}
to obtain the following results,
  \begin{align}
& p_\rho H_{\mu\nu}^{(1)L,\rho}(\hk_1,\hk_2) = \frac{H_{\mu\nu}^{(0)}(\hk_1)}{x_2-\chi-i\eps}, \\[4pt]
& p_\rho H_{\mu\nu}^{(1)R,\rho}(\hk_1,\hk_2) = \frac{H_{\mu\nu}^{(0)}(\hk_2)}{x_1-\chi+i\eps}, \\[4pt]
& p_\rho H_{\mu\nu}^{(2)L,\rho\sigma}(\hk_1,\hk,\hk_2) = \frac{H_{\mu\nu}^{(1)L,\sigma}(\hk,\hk_2)}{x_2-\chi-i\eps}+\frac{H_{\mu\nu}^{(1)L,\sigma}(\hk_1,\hk_2)}{x-x_2+i\eps}, \\[4pt]
& p_\rho H_{\mu\nu}^{(2)M,\rho\sigma}(\hk_1,\hk,\hk_2) = \frac{H_{\mu\nu}^{(1)R,\sigma}(\hk,\hk_1)}{x_2-\chi-i\eps},\\[4pt]
& p_\rho H_{\mu\nu}^{(2)R,\rho\sigma}(\hk_1,\hk,\hk_2) = \frac{H_{\mu\nu}^{(1)R,\sigma}(\hk_1,\hk_2)}{x-\chi+i\eps}, \\[4pt]
& p_\sigma H_{\mu\nu}^{(2)L,\rho\sigma}(\hk_1,\hk,\hk_2) = \frac{H_{\mu\nu}^{(1)L,\rho}(\hk_1,\hk_2)}{x-\chi-i\eps}, \\[4pt]
& p_\sigma H_{\mu\nu}^{(2)M,\rho\sigma}(\hk_1,\hk,\hk_2) = \frac{H_{\mu\nu}^{(1)L,\rho}(\hk,\hk_2)}{x_1-\chi+i\eps}, \\[4pt]
& p_\sigma H_{\mu\nu}^{(2)R,\rho\sigma}(\hk_1,\hk,\hk_2) = \frac{H_{\mu\nu}^{(1)R,\rho}(\hk_1,\hk)}{x-x_1-i\eps}+\frac{H_{\mu\nu}^{(1)R,\rho}(\hk,\hk_2)}{x_1-x+i\eps}, \\[4pt]
& p_\rho p_\sigma H_{\mu\nu}^{(2)L,\rho\sigma}(\hk_1,\hk,\hk_2) = \frac{H_{\mu\nu}^{(0)}(\hk_1)}{(x-\chi-i\eps)(x_2-\chi-i\eps)}, \\[4pt]
& p_\rho p_\sigma H_{\mu\nu}^{(2)M,\rho\sigma}(\hk_1,\hk,\hk_2) = \frac{H_{\mu\nu}^{(0)}(\hk)}{(x_2-\chi-i\eps)(x_1-\chi+i\eps)}, \\[4pt]
& p_\rho p_\sigma H_{\mu\nu}^{(2)R,\rho\sigma}(\hk_1,\hk,\hk_2) = \frac{H_{\mu\nu}^{(0)}(\hk_2)}{(x-\chi+i\eps)(x_1-\chi+i\eps)}. 
\end{align}


With the simplified hard-scattering tensors,  we obtain the simplified hadronic tensors. For $W_{\mu\nu}^{(1)}$, we have 
\begin{align}
W_{\mu\nu}^{(1)} &=
\int_{k_1,k_2}\Bigg\{\Tr\Big[\frac{n\cdot\phi^{(1)}(k_1,k_2)}{p\cdot n}\cdot\frac{H_{\mu\nu}^{(0)}(\hk_1)}{x_2-\chi-i\eps}\Big] \notag \\[2pt]
&\quad\quad\quad +\Tr\Big[\frac{\phi^{(1)}(k_1,k_2)\cdot n}{p\cdot n}\cdot\frac{H_{\mu\nu}^{(0)}(\hk_2)}{x_1-\chi+i\eps}\Big]\Bigg\} \notag\\[2pt]
&+\int_{k_1,k_2}\sum_{i=1}^{2}\Tr\big[\phi_{\rho'}^{(1)}(k_1,k_2) H_{\mu\nu}^{(1)i,\rho}(\hk_1,\hk_2)\big]\omega_\rho^{\rho'} \notag\\[2pt]
&-\int_{k_1,k,k_2}\Bigg\{ \Tr\Big[\frac{n\cdot\phi_{1\sigma'}^{(1)}(k_1,k,k_2)}{p\cdot n}\cdot\frac{H_{\mu\nu}^{(1)L,\sigma}(\hk,\hk_2)}{x_2-\chi-i\eps}\Big]\omega_\sigma^{\sigma'} \notag\\[2pt]
&\quad  +\Tr\Big[\frac{n\cdot\phi_{1\sigma'}^{(1)}(k_1,k,k_2)}{p\cdot n}\cdot\frac{H_{\mu\nu}^{(1)L,\sigma}(\hk_1,\hk_2)}{x-x_2+i\eps}\Big]\omega_\sigma^{\sigma'} \notag\\[2pt]
&\quad  +\Tr\Big[\frac{n\cdot\phi_{1\sigma'}^{(1)}(k_1,k,k_2)}{p\cdot n}\cdot\frac{H_{\mu\nu}^{(1)R,\sigma}(\hk_2,\hk_1)}{x_2-\chi-i\eps}\Big]\omega_\sigma^{\sigma'} \notag\\[2pt]
&\quad  +\Tr\Big[\frac{n\cdot\phi_{1\sigma'}^{(1)}(k_1,k,k_2)}{p\cdot n}\cdot\frac{H_{\mu\nu}^{(1)R,\sigma}(\hk_1,\hk_2)}{x-\chi+i\eps}\Big]\omega_\sigma^{\sigma'} \notag\\[2pt]
&\quad  +\Tr\Big[\frac{n\cdot\phi_{2\sigma'}^{(1)}(k_1,k,k_2)}{p\cdot n}\cdot\frac{H_{\mu\nu}^{(1)L,\sigma}(\hk_1,\hk_2)}{x-\chi-i\eps}\Big]\omega_\sigma^{\sigma'} \notag\\[2pt]
&\quad  +\Tr\Big[\frac{n\cdot\phi_{2\sigma'}^{(1)}(k_1,k,k_2)}{p\cdot n}\cdot\frac{H_{\mu\nu}^{(1)L,\sigma}(\hk,\hk_2)}{x_1-\chi+i\eps}\Big]\omega_\sigma^{\sigma'} \notag\\[2pt]
&\quad  +\Tr\Big[\frac{n\cdot\phi_{2\sigma'}^{(1)}(k_1,k,k_2)}{p\cdot n}\cdot\frac{H_{\mu\nu}^{(1)R,\sigma}(\hk_1,\hk_2)}{x-x_1-i\eps}\Big]\omega_\sigma^{\sigma'} \notag\\[2pt]
&\quad  +\Tr\Big[\frac{n\cdot\phi_{2\sigma'}^{(1)}(k_1,k,k_2)}{p\cdot n}\cdot\frac{H_{\mu\nu}^{(1)R,\sigma}(\hk,\hk_2)}{x_1-x+i\eps}\Big]\omega_\sigma^{\sigma'}\Bigg\} \notag\\[2pt]
&-\int_{k_1,k,k_2}\sum_{i=1}^{3}\Big\{\Tr\big[\phi_{1\rho'\sigma'}^{(1)}(k_1,k,k_2)H_{\mu\nu}^{(2)i,\rho\sigma}(\hk_1,\hk,\hk_2)\big] \notag\\[2pt]
&\quad  +\Tr\big[\phi_{2\rho'\sigma'}^{(1)}(k_1,k,k_2)H_{\mu\nu}^{(2)i,\rho\sigma}(\hk_1,\hk,\hk_2)\big]\Big\}\omega_\rho^{\rho'}\omega_\sigma^{\sigma'}.  \label{f:hadron1h}
\end{align}
For  $W_{\mu\nu}^{(2)}$, we have 
\begin{align}
W_{\mu\nu}^{(2)} &= \int_{k_1,k,k_2} \Bigg\{ 
\sum_{i=1}^{3} \mathrm{Tr}\Big[\phi^{(2)}_{\rho'\sigma'}(k_1,k,k_2)  H_{\mu\nu}^{(2)i,\rho\sigma}(\hat{k}_1,\hat{k},\hat{k}_2)\Big] \omega^{\rho'}_{\rho} \omega^{\sigma'}_{\sigma} \notag\\[4pt]
& + \mathrm{Tr}\Big[\frac{n^{\rho'}n^{\sigma'}}{(p\cdot n)^2} \phi^{(2)}_{\rho'\sigma'}(k_1,k,k_2)\Big]\cdot
\frac{H_{\mu\nu}^{(0)}(\hat{k}_1)}{(x-\chi-i\epsilon)(x_2-\chi-i\epsilon)} \notag\\[4pt]
&+ \mathrm{Tr}\Big[\frac{n^{\rho'}n^{\sigma'}}{(p\cdot n)^2} \phi^{(2)}_{\rho'\sigma'}(k_1,k,k_2)\Big]\cdot
\frac{H_{\mu\nu}^{(0)}(\hat{k})}{(x_2-\chi-i\epsilon)(x_1-\chi+i\epsilon)} \notag\\[4pt]
& + \mathrm{Tr}\Big[\frac{n^{\rho'}n^{\sigma'}}{(p\cdot n)^2} \phi^{(2)}_{\rho'\sigma'}(k_1,k_1,k_2)\Big]\cdot
\frac{H_{\mu\nu}^{(0)}(\hat{k}_2)}{(x-\chi+i\epsilon)(x_1-\chi-i\epsilon)} \notag\\[4pt]
& + \mathrm{Tr}\Big[\frac{n^{\sigma'}}{p\cdot n} \phi^{(2)}_{\rho'\sigma'}(k_1,k,k_2)\Big]\cdot
\frac{H_{\mu\nu}^{(1)L,\rho}(\hat{k}_1,\hat{k}_2)}{x-\chi-i\epsilon}   \omega^{\rho'}_{\rho} \notag\\[4pt]
& + \mathrm{Tr}\Big[\frac{n^{\sigma'}}{p\cdot n} \phi^{(2)}_{\rho'\sigma'}(k,k,k_2)\Big]\cdot
\frac{H_{\mu\nu}^{(1)L,\rho}(\hat{k},\hat{k}_2)}{x_1-\chi+i\epsilon}   \omega^{\rho'}_{\rho} \notag\\[4pt]
& + \mathrm{Tr}\Big[\frac{n^{\sigma'}}{p\cdot n} \phi^{(2)}_{\rho'\sigma'}(k_2,k_1,k)\Big]\cdot
\frac{H_{\mu\nu}^{(1)R,\rho}(\hat{k}_1,\hat{k}_2)}{x-x_1-i\epsilon}   \omega^{\rho'}_{\rho} \notag\\[4pt]
& + \mathrm{Tr}\Big[\frac{n^{\sigma'}}{p\cdot n} \phi^{(2)}_{\rho'\sigma'}(k,k,k_2)\Big]\cdot
\frac{H_{\mu\nu}^{(1)R,\rho}(\hat{k},\hat{k}_2)}{x_1-x+i\epsilon}   \omega^{\rho'}_{\rho} \notag\\[4pt]
& + \mathrm{Tr}\Big[\frac{n^{\rho'}}{p\cdot n} \phi^{(2)}_{\rho'\sigma'}(k_1,k,k_2)\Big]\cdot
\frac{H_{\mu\nu}^{(1)L,\sigma}(\hat{k}_2,\hat{k}_1)}{x_2-\chi-i\epsilon}   \omega^{\sigma'}_{\sigma} \notag\\[4pt]
& + \mathrm{Tr}\Big[\frac{n^{\rho'}}{p\cdot n} \phi^{(2)}_{\rho'\sigma'}(k_1,k,k_2)\Big]\cdot
\frac{H_{\mu\nu}^{(1)L,\sigma}(\hat{k}_1,\hat{k}_2)}{x-\chi+i\epsilon}   \omega^{\sigma'}_{\sigma} \notag\\[4pt]
& + \mathrm{Tr}\Big[\frac{n^{\rho'}}{p\cdot n} \phi^{(2)}_{\rho'\sigma'}(k_1,k,k_2)\Big]\cdot
\frac{H_{\mu\nu}^{(1)R,\sigma}(\hat{k}_2,\hat{k}_1)}{x_2-\chi-i\epsilon}   \omega^{\sigma'}_{\sigma} \notag\\[4pt]
& + \mathrm{Tr}\Big[\frac{n^{\rho'}}{p\cdot n} \phi^{(2)}_{\rho'\sigma'}(k_1,k,k_2)\Big]\cdot
\frac{H_{\mu\nu}^{(1)R,\sigma}(\hat{k}_1,\hat{k}_2)}{x-\chi+i\epsilon}   \omega^{\sigma'}_{\sigma} \Bigg\}.  \label{f:hadorn2h}
\end{align}

\subsection{Gauge invariant hadronic tensors}

The above expression can be further simplified by performing the integrations. First, integrating over the denominator factors, e.g., $\frac{1}{x_2-x-i\epsilon}$ , and then over the gluon field. Here we only show the resulting hadronic tensors, and leave the derivation details in Appendix \ref{app:hadron}. Finally, we have 
\begin{widetext}
\begin{align}
W_{\mu\nu}^{(0)}+W_{\mu\nu}^{(1)}+W_{\mu\nu}^{(2)}
&= \int_{k_1}d^4z_1 e^{ik_1z_1}\Tr\Big[\bra\bar\psi(0)\big[\Lcal^{(0)}(0;z_1)+\Lcal^{(1)}(0;z_1)+\Lcal^{(2)}(0;z_1)\big]\psi(z_1)\ket H_{\mu\nu}^{(0)}(\hk_1)\Big] \notag\\
&+\int_{k_1,k_2}d^4z_1 d^4z_2 e^{ik_1(z_1-z_2)}e^{ik_2z_2}\sum_{i=1}^{2}\Tr\Big[\Big\{\bra\bar\psi(0)\Lcal^{(0)}(0;z_2)D_\rho\Lcal^{(0)}(z_2;z_1)\psi(z_1)\ket \notag\\
&\quad+\bra\bar\psi(0)\Lcal^{(1)}(0;z_2)D_{\rho'}\Lcal^{(0)}(z_2;z_1)\psi(z_1)\ket
+\bra\bar\psi(0)\Lcal^{(0)}(0;z_2)D_{\rho'}\Lcal^{(1)}(z_2;z_1)\psi(z_1)\ket\Big\}H_{\mu\nu}^{(1)i,\rho}(\hk_1,\hk_2)\Big]\omega_\rho^{\rho'} \notag\\
&+\int_{k_1,k,k_2}d^4z_1 d^4z d^4z_2 e^{ik_1(z_1-z)}e^{ik(z-z_2)}e^{ik_2z_2} \notag\\
&\quad \times  \sum_{i=1}^{3}\Tr\Big[\bra\bar\psi(0)\Lcal^{(0)}(0;z_2)D_\rho\Lcal^{(0)}(z_2;z)D_{\rho'}\Lcal^{(0)}(z;z_1)\psi(z_1)\ket H_{\mu\nu}^{(2)i,\rho\sigma}(\hk_1,\hk,\hk_2)\Big] \omega_\rho^{\rho'}\omega_\sigma^{\sigma'}.  \label{f:sumh}
\end{align}
\end{widetext}
We can rewrite Eq. \eqref{f:sumh} in the usual form as a convolution of the correlation functions and the hard-scattering tensors.
\begin{align}
W_{\mu\nu} & = W_{\mu\nu}^{(0)}+W_{\mu\nu}^{(1)}+W_{\mu\nu}^{(2)} \nonumber \\[2pt]
&= \int\frac{d^4k_1}{(2\pi)^4}\Tr\big[\phi^{(0)}(k_1,p)H_{\mu\nu}^{(0)}(\hk_1)\big] \notag\\[2pt]
&+\int\frac{d^4k_1}{(2\pi)^4}\frac{d^4k_2}{(2\pi)^4}\sum_{i=1}^{2}\Tr\big[\phi_{\rho'}^{(1)}(k_1,k_2)H_{\mu\nu}^{(1)i,\rho}(\hk_1,\hk_2)\big]\omega_\rho^{\rho'} \notag\\[2pt]
&+\int\frac{d^4k_1}{(2\pi)^4}\frac{d^4k}{(2\pi)^4}\frac{d^4k_2}{(2\pi)^4}\sum_{i=1}^{3}\Tr\big[\phi_{\rho'\sigma'}^{(2)}(k_1,k,k_2)\nonumber \\[2pt]
& \qquad \qquad  \times H_{\mu\nu}^{(2)i,\rho\sigma}(\hk_1,\hk,\hk_2)\big]\omega_\rho^{\rho'}\omega_\sigma^{\sigma'}. \label{f:wmunut}
\end{align}
The gauge invariant correlation functions are given by
\begin{align}
\phi^{(0)}(k) &= \int d^4z_1 e^{ikz_1}\bra\bar\psi(0)\Lcal(0;z_1)\psi(z_1)\ket,  \\[2pt]
\phi_\rho^{(1)}(k_1,k_2) &= \int d^4z_1 d^4z_2 e^{ik_1(z_1-z_2)}e^{ik_2z_2}\bra\bar\psi(0)\Lcal(0;z_2)\nonumber \\[2pt] 
& \quad  \times   D_\rho(z_2) \Lcal(z_2;z_1)\psi(z_1)\ket,  \\[2pt]
\phi_{\rho\sigma}^{(2)}(k_1,k,k_2) &= \int d^4z_1 d^4z d^4z_2 e^{ik_1(z_1-z)}e^{ik(z-z_2)}e^{ik_2z_2} \notag\\[2pt]
&\quad \times \bra\bar\psi(0)\Lcal(0;z_2)D_\rho(z_2) \Lcal(z_2;z)\nonumber \\[2pt] 
& \quad   \times  D_\sigma(z) \Lcal(z;z_1)\psi(z_1)\ket, 
\end{align}
where the covariant derivative is defined as  $D_\rho(z) = -i\partial_\rho + A_\rho(z)$. The hard-scattering tensors are given in the following forms,
\begin{align}
&H_{\mu\nu}^{(0)}(\hk) = \gamma_\mu(\slashed{\hk}+\slashed q+m)\gamma_\nu\cdot\frac{2\pi}{2p\cdot q}\delta(x-\chi). \\[2pt]
&H_{\mu\nu}^{(1)L,\rho}(\hk_1,\hk_2) = \gamma_\mu\frac{\slashed{\hk}_2+\slashed q+m}{x_2-\chi-i\eps}\gamma^\rho(\slashed{\hk}_1+\slashed q+m)\gamma_\nu\nonumber \\[2pt] 
& \qquad \qquad   \times   \frac{2\pi}{(2p\cdot q)^2}\delta(x_1-\chi). \\[2pt]
&H_{\mu\nu}^{(1)R,\rho}(\hk_1,\hk_2) = \gamma_\mu(\slashed{\hk}_2+\slashed q+m)\gamma^\rho\frac{\slashed{\hk}_1+\slashed q+m}{x_1-\chi+i\eps}\gamma_\nu\nonumber \\[2pt] 
& \qquad \qquad   \times   \frac{2\pi}{(2p\cdot q)^2}\delta(x_2-\chi). \\[2pt]
&H_{\mu\nu}^{(2)L,\rho\sigma}(\hk_1,\hk,\hk_2) = \gamma_\mu\frac{\slashed{\hk}_2+\slashed q+m}{x_2-\chi-i\eps}\gamma^\rho\frac{\slashed{\hk}+\slashed q+m}{x-\chi-i\eps}\gamma^\sigma \nonumber \\[2pt] 
&  \qquad \qquad   \times   (\slashed{\hk}_1+\slashed q+m)\gamma_\nu\frac{2\pi}{(2p\cdot q)^3}\delta(x_1-\chi). \\[2pt]
&H_{\mu\nu}^{(2)M,\rho\sigma}(\hk_1,\hk,\hk_2) = \gamma_\mu\frac{\slashed{\hk}_2+\slashed q+m}{x_2-\chi-i\eps}\gamma^\rho(\slashed{\hk}+\slashed q+m)\gamma^\sigma   \nonumber \\[2pt] 
& \qquad \qquad   \times   \frac{\slashed{\hk}_1+\slashed q+m}{x_1-\chi+i\eps}\gamma_\nu\frac{2\pi}{(2p\cdot q)^3}\delta(x-\chi). \\[2pt]
&H_{\mu\nu}^{(2)R,\rho\sigma}(\hk_1,\hk,\hk_2) = \gamma_\mu(\slashed{\hk}_2+\slashed q+m)\gamma^\rho\frac{\slashed{\hk}+\slashed q+m}{x-\chi+i\eps}\gamma^\sigma   \nonumber \\[2pt] 
& \qquad \qquad   \times   \frac{\slashed{\hk}_1+\slashed q+m}{x_1-\chi+i\eps}\gamma_\nu\frac{2\pi}{(2p\cdot q)^3}\delta(x_2-\chi). 
\end{align}

To further simplify the hadronic tensor, we can integrate the momentum fractions in the hard-scattering tensors. Under this case, we obtain the final results. To be explicit, we write $W_{\mu\nu}^{(0)}$, $W_{\mu\nu}^{(1)}$, and $W_{\mu\nu}^{(2)}$ one by one. 
\paragraph{$W_{\mu\nu}^{(0)}(g=0)$ }
\begin{align}
W_{\mu\nu}^{(0)} &= \frac{1}{2p^+}\Tr\big[\hat\phi^{(0)}(x) \hat h_{\mu\nu}^{(0)}\big],   \label{f:hard0f} \\[2pt]
\hat\phi^{(0)}(\chi) &= p^+\int dz^- e^{i\chi p^+z^-}\bra\bar\psi(0)\Lcal(0;z_1)\psi(z_1)\ket,  \label{f:phi0f}\\[2pt]
\hat h_{\mu\nu}^{(0)} &= \gamma^\mu\slashed n\gamma^\nu + \frac{m}{q^-}\gamma^\mu \gamma^\nu.   \label{f:hardsca0f} 
\end{align}

\paragraph{$W_{\mu\nu}^{(1)}(g=1)$ }
\begin{align}
W_{\mu\nu}^{(1)} &= \frac{1}{4p\cdot q}\Tr\big[\hat\phi_\rho^{(1)}(x) \hat h_{\mu\nu}^{(1)L,\rho}\big].   \label{f:hard1f}  \\[2pt]
\hat\phi_\rho^{(1)}(\chi) &= p^+\int dz^- e^{i\chi p^+z^-}\bra\bar\psi(0)D_{T\rho}(0)\Lcal(0;z_1)\psi(z_1)\ket. \\[2pt]
\hat h_{\mu\nu}^{(1)L,\rho} &= \gamma^\mu\slashed n\gamma^\rho\slashed n\gamma^\nu + \frac{m}{q^-}\gamma_\mu\slashed n\gamma^\rho \gamma_\nu.  \label{f:hardsca1f} 
\end{align}

\paragraph{$W_{\mu\nu}^{(2)M}(g=2)$ }
\begin{align}
W_{\mu\nu}^{(2)M} &= \frac{1}{(2p\cdot q)^2}\Tr\big[\hat\phi_{\rho\sigma}^{(2)M}(x) \hat h_{\mu\nu}^{(2)M,\rho\sigma}\big].    \label{f:hard2mf} \\[2pt]
\hat\phi_{\rho\sigma}^{(2)M}(\chi) &= p^+\int dz^- e^{i\chi p^+z^-}\bra\bar\psi(0)D_{T\rho}(0)\nonumber\\[2pt]
&\qquad  \times \Lcal(0;z)D_{T\sigma}(z)\psi(z)\ket. \\[2pt]
\hat h_{\mu\nu}^{(2)M,\rho\sigma} &= \gamma_\mu\gamma^\rho\slashed p\gamma^\sigma\gamma_\nu+\cdots 
\end{align}

\paragraph{$W_{\mu\nu}^{(2)L}(g=2)$ }
\begin{align}
W_{\mu\nu}^{(2)L} &= \frac{1}{(2p\cdot q)^2}\Tr\big[\hat\phi_{\rho\sigma}^{(2)L}(x) \hat h_{\mu\nu}^{(2)L,\rho\sigma}\big].    \label{f:hard2lf} \\[2pt]
\hat\phi_{\rho\sigma}^{(2)L}(\chi) &= p^+\int dz^- e^{i\chi p^+z^-}\int_0^\infty p^+ dy \bra\bar\psi(0)\Lcal(0;y)  \nonumber\\[2pt]
&\qquad \times  D_{T\rho}(y) D_{T\sigma}(y) \Lcal(y;z)\psi(z)\ket. \\[2pt]
\hat h_{\mu\nu}^{(2)L,\rho\sigma} &= q^-\gamma_\mu\gamma^\rho\slashed n\gamma^\sigma\gamma_\nu+\cdots. \label{f:corre2l}
\end{align}
We can see that all the correlation functions appearing in these expressions are gauge invariant due to the appearance of  gauge links.  We also note that contributions beyond twist-4 have been omitted.


\section{Discussion and Application}\label{sec:applica}

\subsection{Twist Analysis and Massless Limit}

Equations \eqref{f:hard0f}-\eqref{f:corre2l} give the final expressions for the hadronic tensor in inclusive DIS up to twist-4 accuracy, and constitute the main results of the massive collinear expansion.  To further clarify the power structure of the massive collinear expansion, we next analyze the twist classification of the obtained hadronic tensor. For a recent review of higher twists in QCD, see Ref. \cite{Melnitchouk:2026isw}. The contributions in these hadronic tensors can be systematically organized according to their power behavior in $1/Q$. In particular, the explicit quark-mass dependence introduces additional power-suppressed terms, since the quark mass is of $\Lambda_{QCD}$ scale,  and therefore $m/Q$ is counted as a twist-3 contribution, see the second term is Eq. \eqref{f:hardsca0f}. Moreover, each correlation function can contain contributions of different twist orders, when decomposed in terms of PDFs. To have an intuitive understanding of the twist classification in inclusive DIS, we summarize the different contributions in Table \ref{tab:twist}. 

Here we rewrite the hard-scattering tensors as 
\begin{align}
\hat h_{\mu\nu}^{(0)} &= \gamma^\mu\slashed n\gamma^\nu + \frac{m}{q^-}\gamma^\mu \gamma^\nu=h^{(0)}_{\mu\nu,0}+h^{(0)}_{\mu\nu,m } ,    \\
\hat h_{\mu\nu}^{(1)L,\rho} &= \gamma^\mu\slashed n\gamma^\rho\slashed n\gamma^\nu + \frac{m}{q^-}\gamma_\mu\slashed n\gamma^\rho \gamma_\nu=h^{(1)}_{\mu\nu,0}+h^{(1)}_{\mu\nu,m },
\end{align}
and systematically decompose the correlation functions as 
\begin{align}
  \phi^{(i)}=\phi^{(i)}_{t2}+\phi^{(i)}_{t3}+\phi^{(i)}_{t4}. 
\end{align}
Superscripts $i=0,1$ denote the quark-quark correlation function and the quark-gluon-quark correlation function, respectively, and subscripts $t2, t3$ and $t4$ denote leading twist, twist-3 and twist-4.

\begin{table}
    \centering \renewcommand\arraystretch{1.5}
    \caption{Twist classification of the contributions to the hadronic tensor. Those terms in the square brackets are quark mass induced ones.} \label{tab:twist}
    \begin{tabular}{cccc}
        \hline\hline
          &  2 &  3 &   4 \\
        \hline 
      ~ ~~ $W_{\mu\nu}^{(0)}$~~~ & ~~~$h^{(0)}_{\mu\nu,0}\otimes \phi^{(0)}_{t2}$~~~&  \makecell{~~~$h^{(0)}_{\mu\nu,0}\otimes \phi^{(0)}_{t3}$~~~ \\[6pt] ~~~$\left[\textcolor[rgb]{1.00,0.00,0.00}{h^{(0)}_{\mu\nu,m}\otimes \phi^{(0)}_{t2}}\right]$~~~}&   \makecell{~~~$h^{(0)}_{\mu\nu,0}\otimes \phi^{(0)}_{t4}$~~~ \\[6pt]  ~~~$\left[\textcolor[rgb]{1.00,0.00,0.00}{h^{(0)}_{\mu\nu,m}\otimes \phi^{(0)}_{t3}}\right]$~~~} \\[20pt] 
        $W_{\mu\nu}^{(1)}$ &    &  $h^{(1)}_{\mu\nu,0}\otimes \phi^{(1)}_{t3}$ &  \makecell{$h^{(1)}_{\mu\nu,0}\otimes \phi^{(1)}_{t4}$ \\[6pt]  $\left[\textcolor[rgb]{1.00,0.00,0.00}{h^{(1)}_{\mu\nu,m}\otimes \phi^{(1)}_{t3}}\right]$}  \\ [20pt]
        $W_{\mu\nu}^{(2)}$ &   &  &  $h^{(2)}_{\mu\nu}\otimes \phi^{(2)}_{t4}$ \\
        \hline \hline
    \end{tabular}
\end{table}

From Eqs. \eqref{f:hardsca0f} and \eqref{f:hardsca1f}, we notice that the quark mass dependence indeed appears explicitly in the final results. An important feature of the massive case is that the longitudinal momentum fraction of the quark is described by the rescaled variable $\chi$, rather than by the usual variable $x$ used in the massless case.  This rescaling arises naturally from the on-shell condition for the massive quark and is consistently incorporated throughout the collinear expansion. However, considering  the massless limit, $m\to 0$, the mass-dependent terms in the hard-scattering tensors  vanish, i.e., $h^{(0)}_{\mu\nu,m } =h^{(1)}_{\mu\nu,m } =0$. At the same time, the rescaled longitudinal momentum fraction $\chi$ reduces to the conventional momentum fraction $x$ ($\xi$ for massive target). Consequently, the gauge-invariant hadronic tensors obtained above  recover the corresponding results of the massless collinear expansion \cite{Liang:2006wp,Wei:2016far}.

\subsection{Transversity}

As mentioned in the Introduction, transversity is a chiral-odd distribution function and does not appear in the inclusive DIS at leading twist level because the electromagnetic interaction conserves chirality in the massless limit \cite{Jaffe:1991kp,Jaffe:1991ra,Jaffe:1997yz}. A nonzero quark mass, however, can provide a chirality-flipping mechanism and appears to allow a transversity contribution at twist-3 \cite{Jaffe:1993xb,Boer:1996ay,Scopetta:1997qg,Jaffe:1997yz}.   In this part, we therefore investigate the quark-mass-induced transversity contribution at twist-3 accuracy.

To calculate the hadronic tensor, we should decompose the gauge invariant correlation functions first. Correlation functions are $4\times 4$ matrices, they can be decomposed in terms of gamma-matrices, e.g., 
\begin{align}
  \hat{\phi}^{(0)}=I \phi^{(0)} +i\gamma^5 \tilde\phi^{(0)}+\gamma^\alpha \phi^{(0)}_\alpha + \gamma^\alpha\gamma^5 \tilde\phi^{(0)} +i\sigma^{\alpha\rho}\gamma^5 \phi^{(0)}_{\alpha\rho}.
\end{align}
The coefficient functions are further decomposed as 
\begin{align}
\phi^{(0)} &= M  e , \\
\tilde\phi^{(0)}  &= \lambda_h M  e_L , \\
\phi_\alpha^{(0)}  &= p^+ \nbar_\alpha  f_1  - M  \varepsilon_{\perp\alpha S_T}  f_T  + \frac{M^2}{p^+}  n_\alpha  f_3 , \label{f:phi0fdem}\\
\tilde\phi_\alpha^{(0)}  &= \lambda_h  p^+ \nbar_\alpha  g_{1L}  - M  S_{T\alpha}  g_T   + \lambda_h \frac{M^2}{p^+}  n_\alpha  g_{3L} , \\
\phi_{\alpha\rho}^{(0)}  &= p^+ \nbar_\alpha S_{T\rho}  h_{1T}  + M  \varepsilon_{\perp\alpha\rho}  h + \lambda_h M  \nbar_\alpha n_\rho  h_L  \notag\\
  &\quad + \frac{M^2}{p^+}  n_\alpha S_{T\rho}  h_{3T},
\end{align}
 where $\lambda_h$ and  $S_T$ denote the helicity and the transverse polarization of the incoming proton, respectively. Here the spin vector is parametrized as 
 \begin{align}
   S^\mu=\lambda_h \frac{p^+ }{M}\nbar^\mu +S_T^\mu -\lambda_h\frac{M}{2p^+} n^\mu.
 \end{align}
 For the quark-gluon-quark correlation function, we have 
\begin{align}
\phi_{\alpha\rho}^{(1)}  &= -p^+ \nbar_\alpha M  \varepsilon_{\perp\rho S_T}  f_{dT}  + M^2  g_{\perp\alpha\rho}  f_{3d}\notag\\[2pt]
  &\quad   + i\lambda_h M^2  \varepsilon_{\perp\alpha\rho}  f_{3dL} ,\\[2pt]
\tilde\phi_{\alpha\rho}^{(1)}  &= i  p^+ \nbar_\alpha M S_{T\rho}  g_{dT}   + i M^2  \varepsilon_{\perp\alpha\rho}  g_{3d}\notag\\[2pt]
  &\quad   + M^2  g_{\perp\rho\alpha}  \lambda_h  g_{3dL} .  \\[2pt]
\phi_{\alpha\rho\sigma}^{(1)}  &= p^+ \nbar_{[\alpha M}  \varepsilon_{\perp\rho]\sigma}  h_d  +  M^2  \nbar_{[\alpha}  g_{\perp\rho]\sigma}  \lambda_h h_{3dL}\notag\\[2pt]
  &\quad   + M^2  \nbar_{[\alpha} n_{\rho]} S_{T\sigma}  h_{3dT} .
\end{align}

Now, we calculate the quark-mass-induced transversity contribution. Considering the mass-term in Eq. \eqref{f:hardsca0f} and using Eq. \eqref{f:hard0f} gives 
\begin{align}
W_{\mu\nu}^{(t3,o)} &= \frac{1}{2p^+}  \text{Tr}\big[i\sigma^{\alpha\beta}\gamma^5  \gamma^\mu \gamma^\nu\big] \frac{m}{q^-}  p^+ \nbar_\alpha S_{T\beta}  h_{1T} \notag\\
&= -\frac{2m}{p\cdot q}i \varepsilon_{\perp[\mu S_T}p_{\nu]} h_{1T}(x)\propto \frac{m}{Q} h_{1T}. \label{f:chioddh}
\end{align}
This is the quark-mass-induced transversity contribution, which arises at twist-3 from the quark-quark correlation function. However, this contribution by itself does not satisfy the gauge-invariance constraint, it does not satisfy the constraint  $q^\mu W_{\mu\nu}^{(t3,o)} = 0$.  Therefore, the quark-quark contribution alone is not gauge invariant, and the corresponding quark-gluon-quark contribution must be included to restore gauge invariance. 

For completeness, we include all contributions up to twist-3 in our calculation. For the leading twist contribution, we use Eq. \eqref{f:hard0f} and obtain 
\begin{align}
W_{\mu\nu}^{t2} &= \frac{1}{2p^+}  \text{Tr}\big[\hat \phi^{(0)}\hat h_{\mu\nu}^{(0)}\big] 
= \frac{1}{2p^+}  \text{Tr}\big[\slashed{\bar{n}} \gamma^\mu  \slashed\gamma  \gamma^\nu\big]  p^+ f_1 \notag\\
&\quad + \frac{1}{2p^+}  \text{Tr}\big[\slashed{\bar{n}} \gamma^5 \gamma^\mu  \slashed\gamma  \gamma^\nu\big]  \lambda_h p^+ g_{1L}\notag\\
&= 2\big[\nbar_\mu n_\nu + \nbar_\nu n_\mu - g_{\mu\nu}\big]  f_1 - 2 i \lambda_h  \varepsilon_{\perp\mu\nu}  g_{1L}. \label{f:wt2}
\end{align}
 
The twist-3 contribution consists of two parts, corresponding to the quark-quark and quark-gluon-quark correlation functions, respectively. We first consider the chiral-even part from the quark-quark correlation function. We use Eq. \eqref{f:hard1f} and obtain 
\begin{align}
W_{\mu\nu}^{(t3,e1)} &= \frac{1}{2p^+}  \text{Tr}\big[\hat \phi_\alpha^{(0)}\hat  h_{\mu\nu}^{(0)\alpha}\big]
= -\frac{M}{2p^+}  \text{Tr}\big[\gamma^\alpha \gamma^\mu  \slashed n  \gamma^\nu\big] \varepsilon_{\perp\alpha S_T} f_T \notag\\
&\quad - \frac{M}{2p^+}  \text{Tr}\big[\slashed S_T \gamma^5 \gamma^\mu  \slashed n  \gamma^\nu\big]  g_T \notag\\
&= -\frac{2M}{p^+}  n_{\{\mu}  \varepsilon_{\perp\nu \}S_T}  f_T - \frac{2M}{p^+}  i  \varepsilon_{\perp[\mu S_T} n_{\nu]}  g_T.  \label{f:chieven2}
\end{align}
Here we have also used notations $A^{\{\mu}B^{\nu\}} \equiv A^\mu B^\nu +A^\nu B^\mu$, and $A^{[\mu}B^{\nu]} \equiv A^\mu B^\nu -A^\nu B^\mu$. For the chiral-even  twist-3 contributions  from the quark-gluon-quark correlation function, we have 
\begin{align}
W_{\mu\nu}^{(t3,e2)} &= \frac{1}{4p\cdot q}  \text{Tr}\big[\hat \phi_\rho^{(1)} \hat h_{\mu\nu}^{(1)L,\rho}\big]\notag\\
& = -\frac{p^+ M}{4 p\cdot q}  \text{Tr}\big[\slashed{\bar{n}}\gamma^\mu  \slashed{\bar{n}}\gamma^\rho  \slashed\gamma  \gamma^\nu\big]  \varepsilon_{\perp\rho S_T}  f_{dT} \notag\\
&\quad + \frac{i p^+ M}{4 p\cdot q}  \text{Tr}\big[\slashed{\bar{n}}\gamma^5 \gamma^\mu  \slashed S_T  \slashed\gamma  \gamma^\nu\big]  g_{dT} \notag\\
&= \frac{2M}{p\cdot q}  p_\mu  \varepsilon_{\perp\nu S_T}  f_{dT} - \frac{2M}{p\cdot q}  p_\mu  \varepsilon_{\perp\nu S_T}  g_{dT}. \label{f:chieven3}
\end{align}
 We can use QCD equation-of-motion to relate $f_{dT}(g_{dT})$ to $f_{T}(g_{T})$ by using the following relations,
\begin{align}
\chi p^+  \phi_\rho^{(0)} &= -n^\alpha \Big(\text{Re} \phi_{\alpha\rho}^{(1)L} + \varepsilon_{\perp\rho}{}^\sigma  \text{Im} \tilde\phi_{\alpha\sigma}^{(1)L}\Big), \\
\chi p^+  \tilde\phi_\rho^{(0)} &= -n^\alpha \Big(\text{Re} \tilde\phi_{\alpha\rho}^{(1)L} + \varepsilon_{\perp\rho}{}^\sigma  \text{Im} \phi_{\alpha\rho}^{(1)L}\Big) \notag\\
&+ m  \phi_\rho^{[-i\sigma^{\rho\sigma}\gamma^5]},
\end{align}
and obtain 
\begin{align}
-\chi \Big[f_T - i\Big(g_T + \frac{m}{M\chi}  h_{1T}\Big)\Big] = f_{dT} - g_{dT}. \label{f:eom}
\end{align}
We emphasize that $\phi_\rho^{[-i\sigma^{\rho\sigma}\gamma^5]}$ corresponds exactly the transversity term in the decomposition of the quark-quark correlation function. Equation \eqref{f:eom} was also obtained in Ref. \cite{Mulders:1995dh,Bacchetta:2006tn}. It reflects the QCD equation-of-motion constraint among the twist-3 parton distributions. An important feature of this relation is the appearance of the quark-mass term ($m h_{1T}/(Mx)$), which is absent in the massless case \cite{Chen:2020ugq}. Thus, although transversity is a twist-2 chiral-odd distribution, a finite quark mass allows it to enter the twist-3 relations through quark-mass corrections. In other words, the mass-induced transversity contribution from the  quark-quark correlation function cannot be treated independently, since it is constrained by the corresponding  quark-gluon-quark correlation function through the QCD equation-of-motion.

 We  substitute Eq. \eqref{f:eom}  into Eq. \eqref{f:chieven3}, and obtain 
 \begin{align}
&\quad  W_{\mu\nu}^{(t3,e2)} + \left(W_{\mu\nu}^{(t3,e2)}\right)^* \nonumber \\
&= -\frac{2M}{p\cdot q}  x  p_{\{\mu}  \varepsilon_{\nu\}S_T}  f_T- \frac{2M}{p\cdot q}  ix  \varepsilon_{\perp[\mu S_T} p_{\nu]}  g_T \notag\\
&\quad + \frac{2 m}{p\cdot q}  i  \varepsilon_{\perp[\mu S_T} p_{\nu]}  h_{1T}. \label{f:chieven3sum}
\end{align}
We sum Eq. \eqref{f:chieven2} and Eq. \eqref{f:chieven3sum} together and obtain the complete chiral-even twist-3 hadronic tensor. 
 \begin{align}
W_{\mu\nu}^{t3,e}=&- \frac{2M}{p\cdot q}  \bar q_{\{\mu}  \varepsilon_{\perp \nu\}S_T}  f_T - \frac{2M}{p\cdot q}  i  \varepsilon_{\perp[\mu S_T}  \bar{q}_{\nu]}  g_T \notag\\
& + \frac{2m}{p\cdot q}  i  \varepsilon_{\perp[\mu S_T}  p_{\nu]}  h_{1T}, \label{f:t3final}
\end{align}
where $\bar{q}^\mu=q^\mu +2\chi p^\mu$.  Of course,  the $h_{1T}(x)$ term is chiral-odd. We therefore sum Eq. \eqref{f:t3final},  Eq. \eqref{f:chioddh} and Eq. \eqref{f:wt2} together and obtain
 \begin{align}
W_{\mu\nu}&= 2\big[\nbar_\mu n_\nu + \nbar_\nu n_\mu - g_{\mu\nu}\big]  f_1 - 2i\lambda_h  \varepsilon_{\perp\mu\nu}  g_{1L} \notag\\
&=- \frac{2M}{p\cdot q}  \bar q_{\{\mu}  \varepsilon_{\perp \nu\}S_T}  f_T - \frac{2M}{p\cdot q}  i  \varepsilon_{\perp[\mu S_T}  \bar{q}_{\nu]}  g_T.   \label{f:hadrondis}
\end{align}
This is the final hadronic tensor of the inclusive DIS up to twist-3 accuracy.   The quark-quark contribution proportional to \(m h_{1T}/Q\) does not satisfy the gauge invariance constraint by itself. When the corresponding quark-gluon-quark contribution is included, the mass-induced transversity terms cancel in the complete twist-3 hadronic tensor. Thus, no net transversity contribution survives at twist-3 accuracy in inclusive DIS in the one-photon-exchange approximation. 

Finally, we write down the cross section of the inclusive DIS at twist-3 accuracy, 
\begin{align}
  \frac{d\sigma}{d\chi dy} &= \frac{2\alpha^2 e_q^2}{\pi sy^2} \bigg\{A(y)f_1 +B(y)\lambda_e \lambda_h g_{1L} \nonumber \\
  & -  |S_T|4\kappa_M \chi\Big[ C(y) \sin \phi_S f_T +D(y)\lambda_e\cos\phi_S g_T\Big]\bigg\},\label{f:sigmaxi}
\end{align}
where $\kappa_M=M/Q$, 
\begin{align}
  A(y) &= 1+(1-y)^2, \\
  B(y)&=-y(2-y), \\
  C(y)&=(2-y)\sqrt{1-y}, \\
  D(y)&=y\sqrt{1-y},
\end{align}
 and $\phi_S$ is the azimuthal angle of $\vec{S}_T$. 

\subsection{Numerical estimates}

We investigate the transverse-spin asymmetries described by
Eq.~\eqref{f:sigmaxi}, with particular emphasis on the effects of
finite target and quark masses.
Since the twist-3 distributions
$g_T$ and $f_T$ are not well constrained by experimental data,
we use the results of the spectator-diquark
model~\cite{Mao:2014fma,Mao:2014aoa} for our numerical estimates.
The collinear distributions are obtained by integrating over
the full quark transverse momentum,
\begin{align}
f_T^q(x)
&=\int d^2\boldsymbol{k}_T\,f_T^q(x,k_T^2),\\
g_T^q(x)
&=\int d^2\boldsymbol{k}_T\,g_T^q(x,k_T^2).
\end{align}
Using the model expression for $f_T^q(x,k_T^2)$ given in Ref.~\cite{Mao:2014aoa}, we find
\begin{equation}
f_T^q(x)=0.
\end{equation}
This result is consistent with time-reversal invariance~\cite{Bacchetta:2006tn}.
Consequently, the $f_T$ term in Eq.~\eqref{f:sigmaxi} does not contribute to inclusive DIS, and we focus on the $g_T$ contribution to the $\cos\phi_S$ asymmetry.

In the spectator-diquark model, the $u$- and $d$-quark collinear distributions are constructed as
\begin{align}
g_T^u(x)
&=c_s^2g_T^{(s)}(x)+c_a^2g_T^{(a)}(x),\\
g_T^d(x)
&=c_{a'}^2g_T^{(a')}(x).
\end{align}
Here, $s$ denotes the scalar $ud$ diquark, while $a$ and $a'$
denote the axial-vector $ud$ and $uu$ diquarks, respectively.
The coefficients $c_s$, $c_a$, and $c_{a'}$ determine the
relative weights of these spectator contributions in the diquark model.

The corresponding integrated scalar- and axial vector- diquark
contributions read
\begin{align}
g_T^{(s)}(x)
&=
\frac{N_s^2(1-x)^2}{96\pi^2(L_s^2)^3}
\biggl[
2(1-x)(m+M)(m+xM)
\nonumber\\
&\hspace{1.5cm}
-\left(x+\frac{m}{M}\right)(2M_s^2+L_s^2)
\biggr],\\
g_T^{(X)}(x)
&=
\frac{N_X^2(1-x)}{96\pi^2(L_X^2)^2}
\left[x(1-x)-(1+x)\frac{m}{M}\right],
\qquad X=a,a',
\end{align}
where
\begin{equation}
L_X^2
=x M_X^2+(1-x)\Lambda_X^2-x(1-x)M^2,
\qquad X=s,a,a'.
\end{equation}
Here, $M
=0.938~\text{GeV}$ and $m=0.3~\text{GeV}$ are the proton and quark masses,
respectively, while $M_X$, $\Lambda_X$, and $N_X$ denote
the diquark masses, cutoff parameters, and normalization
constants for each spectator channel.
The coefficients $c_s$, $c_a$, and $c_{a'}$, together with
the model parameters $M_X$, $\Lambda_X$, and $N_X$,
are adopted from Ref.~\cite{Bacchetta:2008af}.
The unpolarized distributions $f_1^q(x)$ are evaluated within
the same spectator-diquark model~\cite{Bacchetta:2008af}.
We retain only the $u$ and $d$ valence-quark contributions
and neglect QCD evolution in these numerical estimates.

For the numerical estimates, we consider JLab and EicC
kinematics.
At JLab, we use a $5.5~\mathrm{GeV}$ electron beam
incident on a proton target and impose
\begin{equation}
0.1<x<0.6,\qquad
Q^2>1~\mathrm{GeV}^2,\qquad
W^2>4~\mathrm{GeV}^2.
\end{equation}
For EicC, we use $\sqrt{s}=16.7~\mathrm{GeV}$ and adopt~\cite{Anderle:2021wcy}
\begin{equation}
\begin{aligned}
&0.005<x<0.5,\qquad 0.07<y<0.9,\\
&Q^2>1~\mathrm{GeV}^2,\qquad
W^2>4~\mathrm{GeV}^2.
\end{aligned}
\end{equation}

At fixed measured Bjorken variable $x$ and squared
center-of-mass energy $s$, we integrate over the allowed
$y$ region, with $Q^2=xy(s-M^2)$ and $W^2=M^2+\frac{1-x}{x}Q^2$.
According to Eqs.~\eqref{f:xi} and~\eqref{f:chib}, the mass-corrected partonic
momentum fraction $\chi=\chi(x,y)$ varies with $y$ during
this integration.
We retain this dependence in the mass-corrected calculation
and present both the corrected and uncorrected predictions
as functions of the same measured variable $x$.

The transverse-spin-dependent $\cos\phi_S$ asymmetry
is defined as
 \begin{equation}
A^{\cos\phi_S}(x)=-4\,
\frac{\displaystyle
\int_{\Omega}\frac{dy}{y^2}\,J(x,y)\frac{M}{Q}\chi D(y)
\sum_q e_q^2g_T^q(\chi)}{\displaystyle\int_{\Omega}\frac{dy}{y^2}\,
J(x,y)A(y)\sum_q e_q^2f_1^q(\chi)}.
\label{asycos}
\end{equation}
Here, $\Omega$ is determined by the experimental kinematic
cuts.
For a differential cross section expressed in $(\chi,y)$,
the Jacobian $J(x,y)=|\left.\frac{\partial\chi}{\partial x}\right|_{y,s}|$
accounts for the conversion to the measured variables $(x,y)$.
The uncorrected prediction is obtained by replacing $\chi$
with $x$ in both the explicit factor and the arguments of
the distribution functions, and setting $J=1$.
Both calculations retain the same measured kinematics,
model parameters, and twist-3 prefactor $M/Q$.

 \begin{figure}
     \centering
\includegraphics[width=\linewidth]{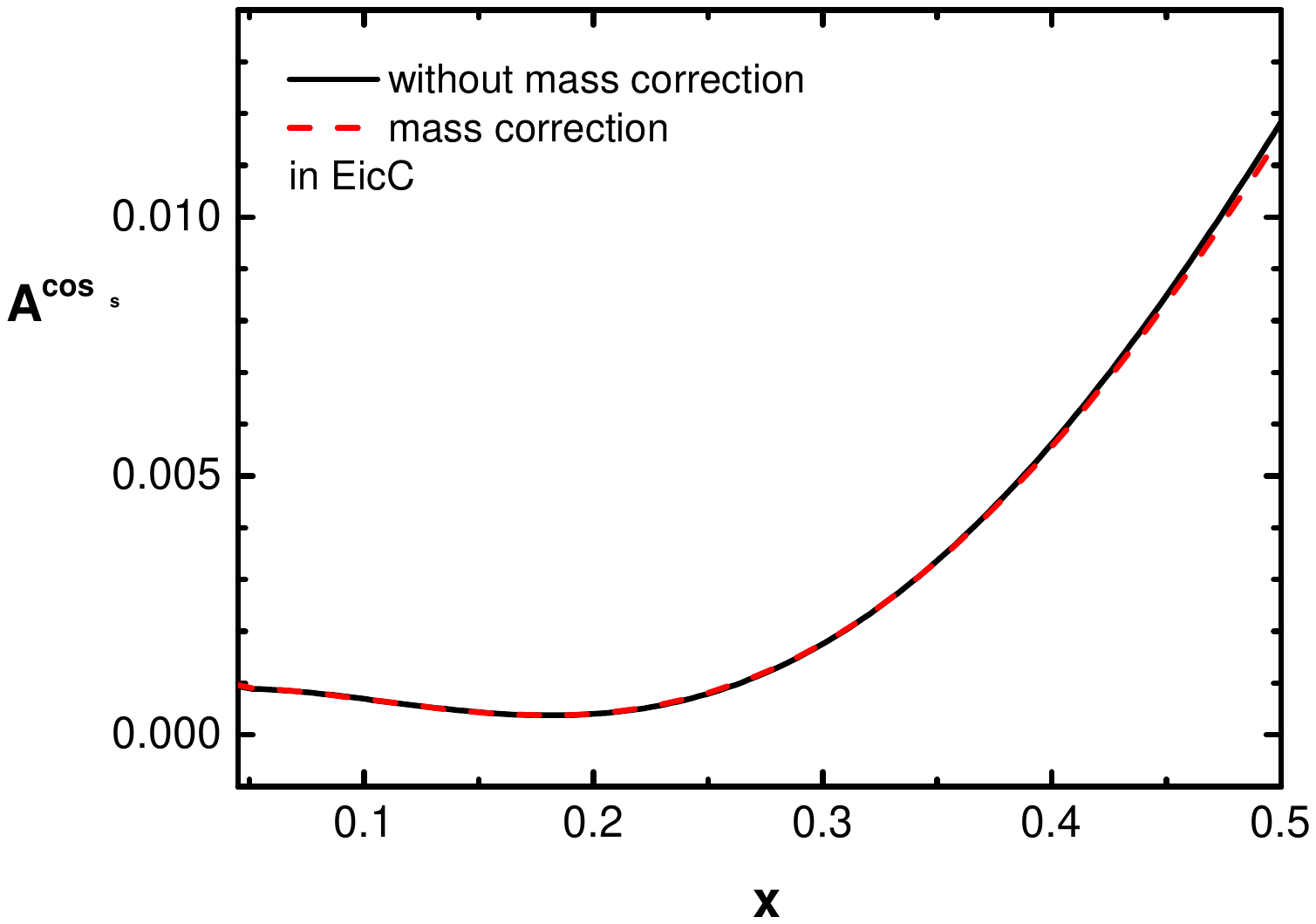}
\includegraphics[width=\linewidth]{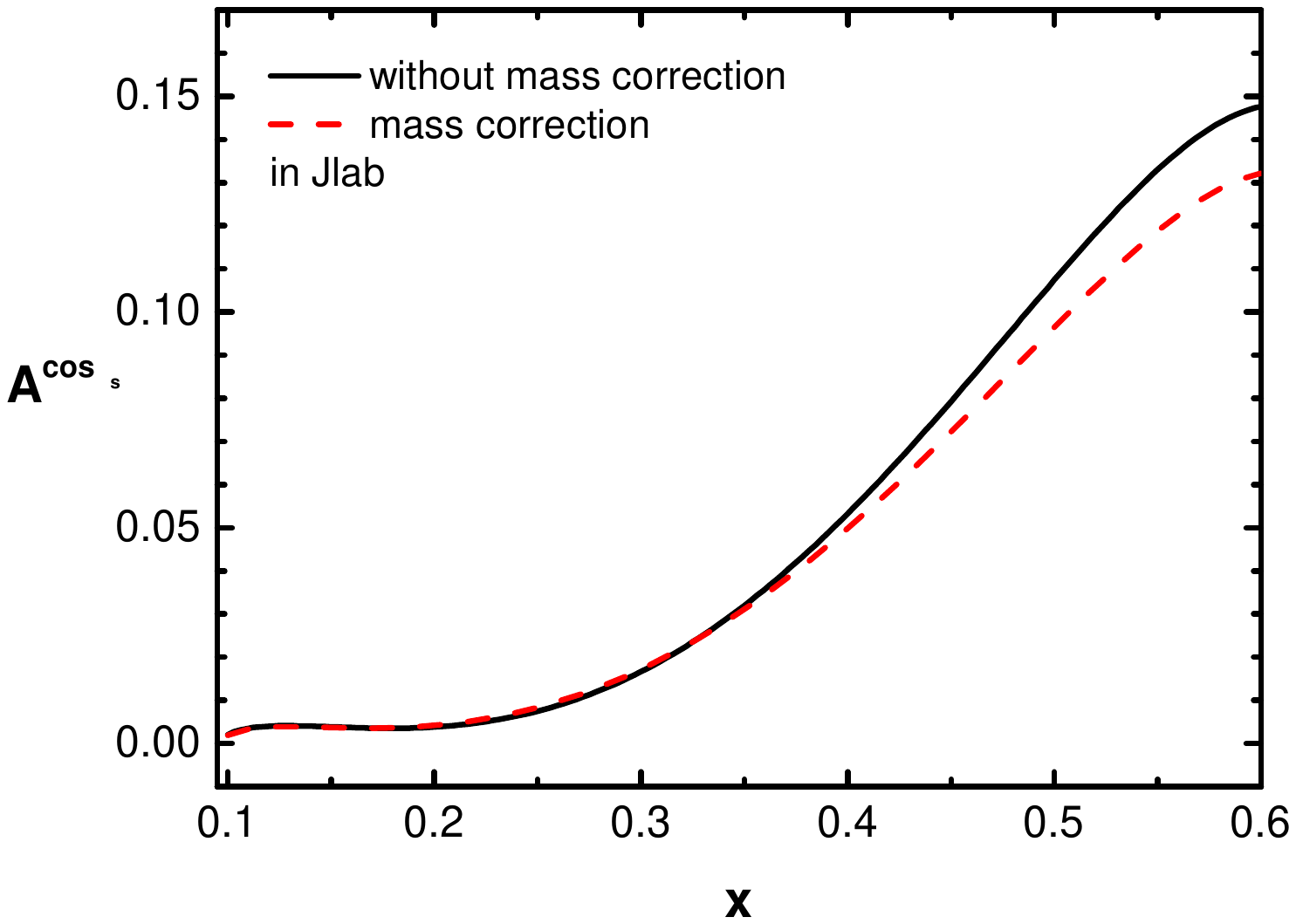}
    \caption{The $x$ dependence of the $g_T$ contribution to the inclusive-DIS
${\cos\phi_S}$ asymmetry in EicC (upper panel) and JLab (lower panel)
kinematics. The solid black curves show the results without mass corrections, while the dashed red curves include target- and quark-mass corrections.}\label{fig:asygt}
 \end{figure}
 
Figure~\ref{fig:asygt} shows the $\cos\phi_S$ asymmetry
in EicC (upper panel) and JLab (lower panel) kinematics.
The solid black curves represent the uncorrected predictions,
while the dashed red curves include mass corrections.
In EicC kinematics, the two curves nearly overlap, indicating
that mass corrections can be largely neglected under the
adopted cuts.
In contrast, mass corrections visibly reduce the asymmetry
at large $x$ in JLab kinematics, where relatively lower
$Q^2$ values are sampled.
This contrast is consistent with enhanced sensitivity to
kinematic mass corrections at lower $Q^2$ and larger $x$.

In particular, the selected EicC kinematics provide a favorable balance: the predicted twist-3 asymmetry retains an appreciable magnitude despite its explicit $M/Q$ suppression, while the mass-dependent $x\rightarrow\chi$ rescaling corrections, controlled by the ratios $M^2x^2/Q^2$ and $m^2/Q^2$, remain small. Within the present model and kinematic cuts, this balance highlights EicC as a particularly promising platform for investigating potentially measurable twist-3 spin asymmetries and probing the parton distribution $g_T$ with limited sensitivity to kinematic mass corrections. These findings also motivate extending the analysis to semi-inclusive processes, where the relevant distribution and fragmentation functions can be investigated.

\section{Summary}\label{sec:sum}

Deep inelastic scattering (DIS) is an important process for investigating the internal structure of the nucleon. The collinear expansion approach has played an important role in these studies. It  can not only provide a systematic framework for organizing contributions according to their twist but also naturally lead to gauge-invariant correlation functions and PDFs. In particular, it allows the hadronic tensor to be expressed in a relatively simple and transparent form up to twist-4 accuracy. However, previous applications of the collinear expansion have generally been formulated in the massless-quark approximation. This treatment is not sufficient when finite quark-mass effects become relevant.  It is therefore necessary to develop a systematic extension of the collinear expansion that consistently incorporates quark-mass effects.
In this paper, we present a systematic formulation of the collinear expansion for inclusive DIS with finite quark masses and derive the hadronic tensor up to twist-4 accuracy. However, our calculation are not complete at twist-4 level. For example, we did not consider contributions from four-quark correlation functions, and only focused on contributions from quark-j-gluon-quark correlation functions. In our results, the quark-mass dependence is retained in the hard-scattering tensors, while the nonperturbative contributions are expressed in terms of gauge-invariant correlation functions. This formulation provides a systematic framework for describing quark-mass-dependent contributions at different twist levels. The resulting hadronic tensor can be organized according to the corresponding power behavior in $1/Q$. In particular, the explicit quark-mass dependence gives rise to additional power-suppressed contributions. Since the quark mass is of order $\Lambda_{\mathrm{QCD}}$, the ratio $m/Q$ can be counted as a twist-3 contribution in the power expansion. However, in the massless limit, $m\to 0$, the mass-dependent contributions vanish and the resulting hadronic tensor reduces to the conventional massless result.

As an application, we investigate the role of transversity in inclusive DIS with a transversely polarized target. Since a finite quark mass provides a chirality-flipping mechanism, the chiral-odd transversity contribution therefore can appear at twist-3. We find that the mass-induced transversity contribution is proportional to $m h_{1T}/Q$, which arises from the quark-quark correlation function. However, the corresponding quark-gluon-quark contribution exactly cancels this term in the complete twist-3 hadronic tensor. In other words, no net transversity contribution remains at twist-3 accuracy in inclusive DIS under the one-photon-exchange approximation. This cancellation demonstrates that the mass-induced transversity term from the quark-quark correlation function cannot be considered as an independent physical contribution, and that the corresponding multiparton correlation is essential for obtaining the complete gauge-invariant result. However, the transversity can contribute to inclusive DIS observables through two-photon-exchange effects \cite{Afanasev:2007ii}. 
For the twist-3 contributions, we numerically estimate the $g_T$-related asymmetry, focusing on the correction effects of finite target and quark masses. We find that the mass corrections become more important
at lower $Q^2$ and larger $x$. In particular, the selected EicC kinematics condition allows the $x\rightarrow\chi$ corrections to be largely neglected, while the predicted asymmetry remains at an appreciable and potentially measurable level despite its explicit $M/Q$ suppression. This favorable balance makes EicC a promising platform for studying twist-3 spin asymmetries and probing the parton distribution $g_T$.

The framework developed in this work provides a systematic starting point for further studies of finite-mass effects in more differential processes. In particular, its extension to jet production in semi-inclusive processes could provide a useful framework for investigating quark-mass effects. Although a collinear expansion formalism for jet production with finite quark masses has yet to be established, previous studies have demonstrated that jet functions can be introduced in inclusive jet production processes to probe chiral-odd effects \cite{Yang:2026cts,Accardi:2017pmi,Accardi:2018gmh,Accardi:2020iqn}. These results suggest a promising avenue for extending the present formalism to finite-mass jet production, potentially providing further insight into the interplay between quark-mass effects and nucleon structure, as well as the higher-twist dynamics of QCD.

\section*{Acknowledgements}
This work was supported by the National Natural Science Foundation of China (Grant Nos. 12405103 and 12305088) and the Natural Science Foundation of Henan Province under Grant Nos. 242300421377.

\begin{appendix}

\section{Derivation of  the gauge-invariant hadronic tensor} \label{app:hadron}
  
From Eqs. \eqref{f:hadron1h} and \eqref{f:hadorn2h}, we integrate over the denominator factors,  and then over the gluon field. After the calculation, we obtain the following expression.
\begin{widetext}
  \begin{align}
  W_{\mu\nu}^{(1)} &=\int_{k_1}\int d^4z_1 e^{ik_1 z_1}\Tr \left\{\bra \bar{\psi}(0)\Big[i\int_{0}^{\infty} A^+(0,y^-,0)dy^- - i\int_{z_1}^{\infty} A^+(z_1^+,y^-,\vec z_{1T})dy^-\Big]\psi(z_1) \ket H^{(0)}_{\mu\nu}(\hk_1)\right\} \nonumber \\
  &+\int_{k_1, k_2} \sum_{i=1}^2 \operatorname{Tr}\left[\phi_{\rho'}^{(1)}(k_1, k_2) H_{\mu \nu}^{(1) i, \rho}(\hat{k}_1, \hat{k}_2)\right] \omega_\rho^{\rho'} \nonumber  \\
  & - \int_{k_1 k_2} d^4 z_1d^4z_2 e^{i k_1(z_1-z_2)} e^{i k_2 z_2} \sum_{i=1}^2 \Tr\Bigg\{\langle p| \bar{\psi}(0)\Big[i \int_0^{\infty} A^+(0,y^-,0) d y^- -i \int_{z_2}^{\infty} A^{+}(z_2,y^-,\vec z_{2T}) dy^-\Big] \nonumber \\
  & \qquad \qquad\qquad \qquad \times i\partial_{\sigma^{\prime}}(z_1)\psi(z_1)\ket H_{\mu \nu}^{(1) i, \sigma}(\hk_1, \hat{k}_2)\Bigg\} \omega_\sigma^{\sigma^{\prime}} \nonumber \\
    & - \int_{k_1 k_2} d^4 z_1d^4z_2 e^{i k_1(z_1-z_2)} e^{i k_2 z_2} \sum_{i=1}^2 \Tr\Bigg\{\langle p| \bar{\psi}(0)\big[ i\partial_{\sigma^{\prime}}(z_2)+ i\partial_{\sigma^{\prime}}(z_1)\big]  \nonumber \\
  & \qquad \qquad\qquad \qquad \times \Big[i \int_{z_2}^{\infty} A^+(z_2,y^-,\vec z_{2T}) d y^- -i \int_{z_1}^{\infty} A^{+}(z_1,y^-,\vec z_{1T}) dy^-\Big]\psi(z_1)\ket H_{\mu \nu}^{(1) i, \sigma}(\hk_1, \hat{k}_2)\Bigg\} \omega_\sigma^{\sigma'}  \nonumber \\
 & -\int_{k_1, k,k_2} \sum_{i=1}^3 \Bigg\{\Tr \Big[\phi_{1\rho'\sigma'}^{(1)}(k_1,k, k_2) H_{\mu \nu}^{(2) i, \rho\sigma}(\hat{k}_1,\hk, \hat{k}_2)\Big]+\Tr \Big[\phi_{2\rho'\sigma'}^{(1)}(k_1,k, k_2) H_{\mu \nu}^{(2) i,\sigma\rho}(\hat{k}_1,\hk, \hat{k}_2)\Big]\Bigg\}\omega_\rho^{\rho'}\omega_\sigma^{\sigma'}.
\end{align}
Similarly, we obtain the hadronic tensor $W_{\mu\nu}^{(2)}$,
\begin{align}
W^{(2)}_{\mu\nu} &= \int_{k_1,k,k_2} \sum_{i=1}^{3}\Tr\Big[\phi^{(2)}_{\rho'\sigma'}(k_1,k,k_2)  H^{(2)i,\sigma\rho}_{\mu\nu}(\hk_1,\hk,\hk_2)\Big]  \omega_\sigma^{\sigma'} \omega_\rho^{\rho'} \notag\\[6pt]
&+ \int_{k_1} dz_1 e^{ik_1 z_1}\Tr\Bigg\{\langle p| \bar\psi(0) \bigg[ i \int_0^\infty  A^+(0,y_1^-,0) dy_1^-\;\cdot\; i \int_{y_1^-}^\infty  A^+(0,y_2^-,0) dy_2^- \notag\\
&\hspace{2.4cm} + i \int_0^\infty  A^+(0,y^-,0) dy^-\;\cdot\;(-i) \int_{z_1^-}^\infty  A^+(z_1^+,y^-,\vec z_{1T}) dy^- \notag\\
&\hspace{2.4cm} + (-i) \int_{z_1^-}^\infty  A^+(z_1^+,y_1^-,\vec z_{1T}) dy_2^-\;\cdot\;(-i) \int_{y_2^-}^\infty  A^+(z_1^+,y_1^-,\vec z_{1T}) dy_1^-  \bigg] \psi(z_1) |p\rangle\; H^{(0)}_{\mu\nu}(\hk_1)\Bigg\} \notag\\[8pt]
& + \int_{k_1,k_2} d^4z_1  d^4z_2\; e^{ik_1(z_1-z_2)}  e^{ik_2 z_2}\;\sum_{i=1}^2\Tr \Bigg\{\langle p| \bar\psi(0)  A_{\rho'}(z_2) \bigg[  i \int_{z_2^-}^\infty  A^+(z_2^+,y^-,\vec z_{2T}) dy^- \notag\\
&\hspace{2.4cm} - i \int_{z_2^-}^\infty  A^+(z_1^+,y^-,\vec z_{1T}) dy^- \bigg] \psi(z_1) |p\rangle\; H^{(1)i,\rho}_{\mu\nu}(\hk_1,\hk_2)\Bigg\} \omega_\rho^{\rho'} \notag\\[8pt]
& + \int_{k_1,k_2} d^4z_1  d^4z_2\; e^{ik_1(z_1-z_2)}  e^{ik_2 z_2}\;\sum_{i=1}^2\Tr \Bigg\{\langle p| \bar\psi(0) \bigg[  i \int_0^\infty  A^+(0,y^-,0) dy^- \notag\\
&\hspace{2.4cm} - i \int_{z_2^-}^\infty  A^+(z_2^+,y^-,\vec z_{2T}) dy^- \bigg]  A_{\rho'}(z) \psi(z_1) |p\rangle\; H^{(1)i,\rho}_{\mu\nu}(\hk_1,\hk_2)\Bigg\}  \omega_\rho^{\rho'}.
\end{align}


To deal with gauge links,  we first introduce the following path integral,  
\begin{align}
\Lcal&(z_1^+,z_1^-, \vec z_{1T};\infty) = P \exp \left[i\int_{z_1^-}^\infty A^+(z_1^+,y^-,\vec z_{1T}) dy^-\right] \nonumber \\
&= 1+i\int_{z_1^-}^\infty  A^+(z_1^+,y_1^-,\vec z_{1T}) dy_1^- + (i)^2 \int_{z_1^-}^\infty   dy_1^- \int_{y_1^-}^\infty   dy_2^- A^+(z_1^+,y_1^-,\vec z_{1T})A^+(z_1^+,y_2^-,\vec z_{1T}) +\cdots.
\end{align}
and its conjugate, 
\begin{align}
& \Lcal^\dag(z_1^+,z_1^-, \vec z_{1T};\infty) = 1-i\int_{z_1^-}^\infty  A^+(z_1^+,y_1^-,\vec z_{1T}) dy_1^-   + (-i)^2 \int_{z_1^-}^\infty   dy_1^- \int_{y_1^-}^\infty   dy_2^- A^+(z_1^+,y_1^-,\vec z_{1T})A^+(z_1^+,y_2^-,\vec z_{1T}) +\cdots. 
\end{align}
Hence, we write the gauge link $\Lcal(z_1;z_2)$ in the following form,
\begin{align}
\Lcal(z_1;z_2) &= \Lcal(z_1;\infty) \Lcal^\dagger(z_2;\infty)= \Lcal^{(0)}(z_1;z_2)+\Lcal^{(1)}(z_1;z_2)+\Lcal^{(2)}(z_1;z_2) +\cdots,
\end{align}
where
\begin{align}
\Lcal^{(0)}(z_1;z_2) &= 1, \\[2pt]
\Lcal^{(1)}(z_1;z_2) &= i\int_{z_1^-}^\infty dy^- A^+(z_1^+,y^-,\vec z_{1T})  - i\int_{z_2^-}^\infty dy^- A^+(z_2^+,y^-,\vec z_{1T}), \\[2pt] 
\Lcal^{(2)}(z_1;z_2)& = i^2\int_{z_1^-}^\infty dy_1^-\int_{y_1^-}^\infty dy_2^- A^+(z_1^+,y_1^-,\vec z_{1T})  A^+(z_1^+,y_2^-,\vec z_{1T})+i\int_{z_1^-}^\infty dy^- A^+(z_1^+,y^-,\vec z_{1T})\nonumber \\
& \times   (-i)\int_{z_2^-}^\infty dy^- A^+(z_2^+,y^-,\vec z_{1T}) +(-i)^2\int_{z_2^-}^\infty dy_2^-\int_{y_2^-}^\infty dy_1^- A^+(z_2^+,y_2^-,\vec z_{1T}) A^+(z_2^+,y_1^-,\vec z_{1T}). 
\end{align}
Using these definitions of the gauge links, we can rewrite the hadronic tensors in the following forms.
\paragraph{$W_{\mu\nu}^{(0)}$ }
\begin{align}
W_{\mu\nu}^{(0)} &= \int_{k_1}d^4z_1 e^{ik_1z_1}\Tr\Big[\bra\bar\psi(0)\Lcal^{(0)}(0;z_1)\psi(z_1)\ket H_{\mu\nu}^{(0)}(\hk_1)\Big] \notag\\
&+\int_{k_1,k_2}d^4z_1 d^4z_2 e^{ik_1(z_1-z_2)}e^{ik_2z_2}\sum_{i=1}^{2}\Tr\Big[\bra\bar\psi(0) \Lcal^{(0)}(0;z_2)(-i\partial_{\rho'})\Lcal^{(0)}(z_2;z_1)\psi(z_1)\ket H_{\mu\nu}^{(1)i,\rho}(\hk_1,\hk_2)\Big]\omega_\rho^{\rho'} \notag\\
&+\int_{k_1,k,k_2}d^4z_1 d^4z d^4z_2 e^{ik_1(z_1-z)}e^{ik(z-z_2)}e^{ik_2z_2}\sum_{i=1}^{3}\Tr\Big[\bra\bar\psi(0) \Lcal^{(0)}(0;z_2)(-i\partial_{\rho'})\Lcal^{(0)}(z_2;z)(-i\partial_{\sigma'})\Lcal^{(0)}(z;z_1)\psi(z_1)\ket \nonumber \\
& \hspace{2.5cm} \times H_{\mu\nu}^{(2)i,\rho\sigma}(\hk_1,\hk,\hk_2)\Big]\omega_\rho^{\rho'}\omega_\sigma^{\sigma'}. 
\end{align}

\paragraph{ $W_{\mu\nu}^{(1)}$}
\begin{align}
W_{\mu\nu}^{(1)} &= \int_{k_1}d^4z_1 e^{ik_1z_1}\Tr\Big[\bra\bar\psi(0)\Lcal^{(1)}(0;z_1)\psi(z_1)\ket H_{\mu\nu}^{(0)}(\hk_1)\Big] \notag\\
&+\int_{k_1,k_2}d^4z_1 d^4z_2 e^{ik_1(z_1-z_2)}e^{ik_2z_2}\sum_{i=1}^{2}\Tr\Big[\Big\{\bra\bar\psi(0)\Lcal^{(0)}(0;z_2)A_\rho(z_2) \Lcal^{(0)}(z_2;z_1)\psi(z_1)\ket \notag\\
&\quad +\bra\bar\psi(0)\Lcal^{(1)}(0;z_2)(-i\partial_{\rho'})\Lcal^{(0)}(z_2;z_1)\psi(z_1)\ket +\bra\bar\psi(0)\Lcal^{(0)}(0;z_2)(-i\partial_{\rho'})\Lcal^{(1)}(z_2;z_1)\psi(z_1)\ket\Big\}H_{\mu\nu}^{(1)i,\rho}(\hk_1,\hk_2)\Big]\omega_\rho^{\rho'} \notag\\
&+\int_{k_1,k,k_2}d^4z_1 d^4z d^4z_2 e^{ik_1(z_1-z)}e^{ik(z-z_2)}e^{ik_2z_2}\sum_{i=1}^{3}\Tr\Big[\Big\{\bra\bar\psi(0)\Lcal^{(0)}(0;z_2)A_\rho(z_2)\Lcal^{(0)}(z_2;z)(-i\partial_{\sigma'}) \Lcal^{(0)}(z;z_1)\psi(z_1)\ket \notag\\
&\quad +\bra\bar\psi(0)\Lcal^{(0)}(0;z_2)(-i\partial_{\rho'})\Lcal^{(1)}(z_2;z)(-i\partial_{\sigma'})\Lcal^{(0)}(z;z_1)\psi(z_1)\ket \notag\\
&\quad+\bra\bar\psi(0)\Lcal^{(0)}(0;z_2)(-i\partial_{\rho'})\Lcal^{(0)}(z_2;z)A_{\sigma'}(z)\Lcal^{(0)}(z;z_1)\psi(z_1)\ket\Big\}H_{\mu\nu}^{(2)i,\rho\sigma}(\hk_1,\hk,\hk_2)\Big] \omega_\rho^{\rho'}\omega_\sigma^{\sigma'}.
\end{align}

\paragraph{$W_{\mu\nu}^{(2)}$ }
\begin{align}
W_{\mu\nu}^{(2)} &= \int_{k_1}d^4z_1 e^{ik_1z_1}\Tr\big[\bra\bar\psi(0)\Lcal^{(2)}(0;z_1)\psi(z_1)\ket H_{\mu\nu}^{(0)}(\hk_1)\big] \notag\\
&+\int_{k_1,k_2}d^4z_1 d^4z_2 e^{ik_1(z_1-z_2)}e^{ik_2z_2}\sum_{i=1}^{2}\Tr\Big[\Big\{\bra\bar\psi(0)\Lcal^{(0)}(0;z_2)A_\rho(z_2)\Lcal^{(1)}(z_2;z_1)\psi(z_1)\ket \notag\\
&\quad +\bra\bar\psi(0)\Lcal^{(1)}(0;z_2)A_\rho(z_2)\Lcal^{(0)}(z_2;z_1)\psi(z_1)\ket\Big\}H_{\mu\nu}^{(1)i,\rho}(\hk_1,\hk_2)\Big]\omega_\rho^{\rho'} \notag\\
&+\int_{k_1,k,k_2}d^4z_1 d^4z d^4z_2 e^{ik_1(z_1-z)}e^{ik(z-z_2)}e^{ik_2z_2}\sum_{i=1}^{3}\Tr\Big[\bra\bar\psi(0)\Lcal^{(0)}(0;z_2)A_\rho(z_2)\Lcal^{(0)}(z_2;z)A_\sigma(z)\Lcal^{(0)}(z;z_1)\psi(z_1)\ket \notag\\
&\hspace{5.5cm}\times  H_{\mu\nu}^{(2)i,\rho\sigma}(\hk_1,\hk,\hk_2)\Big]\omega_\rho^{\rho'}\omega_\sigma^{\sigma'}. 
\end{align}
Finally, by summing $W_{\mu\nu}^{(0)}$, $W_{\mu\nu}^{(1)}$, and $W_{\mu\nu}^{(2)}$, we will obtain the final result given in Eq. \eqref{f:sumh}.  

\end{widetext}
  
\end{appendix}

\newpage
\bibliographystyle{apsrev4-2}   

\bibliography{references}

@article{Donnelly:1975ze,
    author = "Donnelly, T. W. and Walecka, J. D.",
    title = "{Electron Scattering and Nuclear Structure}",
    reportNumber = "ITP-485-STANFORD",
    doi = "10.1146/annurev.ns.25.120175.001553",
    journal = "Ann. Rev. Nucl. Part. Sci.",
    volume = "25",
    pages = "329--405",
    year = "1975"
}

@article{Bloom:1969kc,
    author = "Bloom, Elliott D. and others",
    title = "{High-Energy Inelastic e p Scattering at 6-Degrees and 10-Degrees}",
    reportNumber = "SLAC-PUB-0642",
    doi = "10.1103/PhysRevLett.23.930",
    journal = "Phys. Rev. Lett.",
    volume = "23",
    pages = "930--934",
    year = "1969"
}

@article{EuropeanMuon:1987isl,
    author = "Ashman, J. and others",
    editor = "Hughes, V. W. and Cavata, C.",
    collaboration = "European Muon",
    title = "{A Measurement of the Spin Asymmetry and Determination of the Structure Function g(1) in Deep Inelastic Muon-Proton Scattering}",
    reportNumber = "CERN-EP-87-230",
    doi = "10.1016/0370-2693(88)91523-7",
    journal = "Phys. Lett. B",
    volume = "206",
    pages = "364",
    year = "1988"
}

@article{EuropeanMuon:1989yki,
    author = "Ashman, J. and others",
    editor = "Hughes, V. W. and Cavata, C.",
    collaboration = "European Muon",
    title = "{An Investigation of the Spin Structure of the Proton in Deep Inelastic Scattering of Polarized Muons on Polarized Protons}",
    reportNumber = "CERN-EP-89-73",
    doi = "10.1016/0550-3213(89)90089-8",
    journal = "Nucl. Phys. B",
    volume = "328",
    pages = "1",
    year = "1989"
}

@article{Accardi:2012qut,
    author = "Accardi, A. and others",
    editor = "Deshpande, A. and Meziani, Z. E. and Qiu, J. W.",
    title = "{Electron Ion Collider: The Next QCD Frontier}: {Understanding the glue that binds us all}",
    eprint = "1212.1701",
    archivePrefix = "arXiv",
    primaryClass = "nucl-ex",
    reportNumber = "BNL-98815-2012-JA, JLAB-PHY-12-1652",
    doi = "10.1140/epja/i2016-16268-9",
    journal = "Eur. Phys. J. A",
    volume = "52",
    number = "9",
    pages = "268",
    year = "2016"
}

@article{AbdulKhalek:2021gbh,
    author = "Abdul Khalek, R. and others",
    title = "{Science Requirements and Detector Concepts for the Electron-Ion Collider}: {EIC Yellow Report}",
    eprint = "2103.05419",
    archivePrefix = "arXiv",
    primaryClass = "physics.ins-det",
    reportNumber = "BNL-220990-2021-FORE, JLAB-PHY-21-3198, LA-UR-21-20953",
    doi = "10.1016/j.nuclphysa.2022.122447",
    journal = "Nucl. Phys. A",
    volume = "1026",
    pages = "122447",
    year = "2022"
}

@article{Anderle:2021wcy,
    author = "Anderle, Daniele P. and others",
    title = "{Electron-ion collider in China}",
    eprint = "2102.09222",
    archivePrefix = "arXiv",
    primaryClass = "nucl-ex",
    reportNumber = "Frontiers of Physics, Volume 16 Issue (6):64701, 2021",
    doi = "10.1007/s11467-021-1062-0",
    journal = "Front. Phys. (Beijing)",
    volume = "16",
    number = "6",
    pages = "64701",
    year = "2021"
}

@article{Collins:1989gx,
    author = "Collins, John C. and Soper, Davison E. and Sterman, George F.",
    title = "{Factorization of Hard Processes in QCD}",
    eprint = "hep-ph/0409313",
    archivePrefix = "arXiv",
    reportNumber = "ITP-SB-89-31",
    doi = "10.1142/9789814503266_0001",
    journal = "Adv. Ser. Direct. High Energy Phys.",
    volume = "5",
    pages = "1--91",
    year = "1989"
}

@article{Wilson:1969zs,
    author = "Wilson, Kenneth G.",
    title = "{Nonlagrangian models of current algebra}",
    doi = "10.1103/PhysRev.179.1499",
    journal = "Phys. Rev.",
    volume = "179",
    pages = "1499--1512",
    year = "1969"
}

@article{Brandt:1970kg,
    author = "Brandt, Richard A. and Preparata, Giuliano",
    title = "{Operator product expansions near the light cone}",
    doi = "10.1016/0550-3213(71)90265-3",
    journal = "Nucl. Phys. B",
    volume = "27",
    pages = "541--567",
    year = "1971"
}

@article{Christ:1972ms,
    author = "Christ, Norman H. and Hasslacher, B. and Mueller, Alfred H.",
    title = "{Light cone behavior of perturbation theory}",
    doi = "10.1103/PhysRevD.6.3543",
    journal = "Phys. Rev. D",
    volume = "6",
    pages = "3543",
    year = "1972"
}

@article{Ellis:1982wd,
    author = "Ellis, R. Keith and Furmanski, W. and Petronzio, R.",
    title = "{Power Corrections to the Parton Model in QCD}",
    reportNumber = "CERN-TH-3254",
    doi = "10.1016/0550-3213(82)90132-8",
    journal = "Nucl. Phys. B",
    volume = "207",
    pages = "1--14",
    year = "1982"
}

@article{Ellis:1982cd,
    author = "Ellis, R. Keith and Furmanski, W. and Petronzio, R.",
    title = "{Unraveling Higher Twists}",
    reportNumber = "CERN-TH-3301",
    doi = "10.1016/0550-3213(83)90597-7",
    journal = "Nucl. Phys. B",
    volume = "212",
    pages = "29",
    year = "1983"
}

@article{Qiu:1990xxa,
    author = "Qiu, Jian-wei and Sterman, George F.",
    title = "{Power corrections in hadronic scattering. 1. Leading 1/Q**2 corrections to the Drell-Yan cross-section}",
    reportNumber = "ITP-SB-90-49",
    doi = "10.1016/0550-3213(91)90503-P",
    journal = "Nucl. Phys. B",
    volume = "353",
    pages = "105--136",
    year = "1991"
}

@article{Qiu:1990xy,
    author = "Qiu, Jian-wei and Sterman, George F.",
    title = "{Power corrections to hadronic scattering. 2. Factorization}",
    reportNumber = "ITP-SB-90-50",
    doi = "10.1016/0550-3213(91)90504-Q",
    journal = "Nucl. Phys. B",
    volume = "353",
    pages = "137--164",
    year = "1991"
}

@article{Qiu:1988dn,
    author = "Qiu, Jian-Wei",
    title = "{Twist Four Contributions to the Parton Structure Functions}",
    reportNumber = "ANL-HEP-PR-88-10",
    doi = "10.1103/PhysRevD.42.30",
    journal = "Phys. Rev. D",
    volume = "42",
    pages = "30--44",
    year = "1990"
}

@article{Liang:2006wp,
    author = "Liang, Zuo-tang and Wang, Xin-Nian",
    title = "{Azimuthal and single spin asymmetry in deep-inelastic lepton-nucleon scattering}",
    eprint = "hep-ph/0609225",
    archivePrefix = "arXiv",
    reportNumber = "LBNL-61449",
    doi = "10.1103/PhysRevD.75.094002",
    journal = "Phys. Rev. D",
    volume = "75",
    pages = "094002",
    year = "2007"
}

@article{Song:2010pf,
    author = "Song, Yu-kun and Gao, Jian-hua and Liang, Zuo-tang and Wang, Xin-Nian",
    title = "{Twist-4 contributions to the azimuthal asymmetry in SIDIS}",
    eprint = "1012.4179",
    archivePrefix = "arXiv",
    primaryClass = "hep-ph",
    doi = "10.1103/PhysRevD.83.054010",
    journal = "Phys. Rev. D",
    volume = "83",
    pages = "054010",
    year = "2011"
}

@article{Song:2013sja,
    author = "Song, Yu-kun and Gao, Jian-hua and Liang, Zuo-tang and Wang, Xin-Nian",
    title = "{Azimuthal asymmetries in semi-inclusive DIS with polarized beam and/or target and their nuclear dependences}",
    eprint = "1308.1159",
    archivePrefix = "arXiv",
    primaryClass = "hep-ph",
    doi = "10.1103/PhysRevD.89.014005",
    journal = "Phys. Rev. D",
    volume = "89",
    number = "1",
    pages = "014005",
    year = "2014"
}

@article{Wei:2013csa,
    author = "Wei, Shu-yi and Song, Yu-kun and Liang, Zuo-tang",
    title = "{Higher twist contribution to fragmentation function in inclusive hadron production in $e^+e^-$ annihilation}",
    eprint = "1309.4191",
    archivePrefix = "arXiv",
    primaryClass = "hep-ph",
    doi = "10.1103/PhysRevD.89.014024",
    journal = "Phys. Rev. D",
    volume = "89",
    number = "1",
    pages = "014024",
    year = "2014"
}

@article{Wei:2014pma,
    author = "Wei, Shu-Yi and Chen, Kai-bao and Song, Yu-kun and Liang, Zuo-tang",
    title = "{Leading and higher twist contributions in semi-inclusive $e^{+}e^{-}$ annihilation at high energies}",
    eprint = "1410.4314",
    archivePrefix = "arXiv",
    primaryClass = "hep-ph",
    doi = "10.1103/PhysRevD.91.034015",
    journal = "Phys. Rev. D",
    volume = "91",
    number = "3",
    pages = "034015",
    year = "2015"
}

@article{Wei:2016far,
    author = "Wei, Shu-yi and Song, Yu-kun and Chen, Kai-bao and Liang, Zuo-tang",
    title = "{Twist-4 contributions to semi-inclusive deeply inelastic scatterings with polarized beam and target}",
    eprint = "1611.08688",
    archivePrefix = "arXiv",
    primaryClass = "hep-ph",
    doi = "10.1103/PhysRevD.95.074017",
    journal = "Phys. Rev. D",
    volume = "95",
    number = "7",
    pages = "074017",
    year = "2017"
}

@article{Yang:2017sxz,
    author = "Yang, Wei-hua and Chen, Kai-bao and Liang, Zuo-tang",
    title = "{Twist-4 contributions to semi-inclusive $e^+e^$ annihilation process}",
    eprint = "1707.00402",
    archivePrefix = "arXiv",
    primaryClass = "hep-ph",
    doi = "10.1103/PhysRevD.96.054016",
    journal = "Phys. Rev. D",
    volume = "96",
    number = "5",
    pages = "054016",
    year = "2017"
}

@article{Chen:2020ugq,
    author = "Chen, Kai-Bao and Yang, Wei-Hua",
    title = "{Parity violating semi-inclusive deeply inelastic scattering at the Electron-Ion Collider}",
    eprint = "2004.01359",
    archivePrefix = "arXiv",
    primaryClass = "hep-ph",
    doi = "10.1103/PhysRevD.101.096017",
    journal = "Phys. Rev. D",
    volume = "101",
    number = "9",
    pages = "096017",
    year = "2020"
}

@article{Yang:2020ezt,
    author = "Yang, Weihua",
    title = "{Twist-4 contributions to $e^+ e^− \to \Omega X$ process}",
    doi = "10.1016/j.nuclphysa.2020.121729",
    journal = "Nucl. Phys. A",
    volume = "997",
    pages = "121729",
    year = "2020"
}

@article{Yang:2020qsk,
    author = "Yang, Weihua",
    title = "{Charged current semi-inclusive deeply inelastic scattering at the Electron-Ion Collider}",
    eprint = "2011.10212",
    archivePrefix = "arXiv",
    primaryClass = "hep-ph",
    doi = "10.1103/PhysRevD.103.016011",
    journal = "Phys. Rev. D",
    volume = "103",
    number = "1",
    pages = "016011",
    year = "2021"
}

@article{Yang:2022mmh,
    author = "Yang, Weihua and Li, Chao",
    title = "{Forward-backward asymmetry in the electron positron annihilation process at twist-4}",
    eprint = "2205.04068",
    archivePrefix = "arXiv",
    primaryClass = "hep-ph",
    doi = "10.1103/PhysRevD.106.036016",
    journal = "Phys. Rev. D",
    volume = "106",
    number = "3",
    pages = "036016",
    year = "2022"
}

@article{Yang:2022knp,
    author = "Yang, Weihua",
    title = "{Hadron pair production in semi-inclusive electron positron annihilation process at twist-4}",
    doi = "10.1140/epjc/s10052-022-10698-y",
    journal = "Eur. Phys. J. C",
    volume = "82",
    number = "8",
    pages = "741",
    year = "2022"
}

@article{Yang:2022xwy,
    author = "Yang, Weihua and Yang, Xinghua",
    title = "{Intrinsic asymmetries in semi-inclusive deeply inelastic scattering at the Electron-Ion Collider}",
    eprint = "2209.01629",
    archivePrefix = "arXiv",
    primaryClass = "hep-ph",
    doi = "10.1103/PhysRevD.106.093003",
    journal = "Phys. Rev. D",
    volume = "106",
    number = "9",
    pages = "093003",
    year = "2022"
}

@article{Ralston:1979ys,
    author = "Ralston, John P. and Soper, Davison E.",
    title = "{Production of Dimuons from High-Energy Polarized Proton Proton Collisions}",
    reportNumber = "OITS-100",
    doi = "10.1016/0550-3213(79)90082-8",
    journal = "Nucl. Phys. B",
    volume = "152",
    pages = "109",
    year = "1979"
}

@article{Baldracchini:1980uq,
    author = "Baldracchini, F. and Craigie, N. S. and Roberto, V. and Socolovsky, M.",
    title = "{A Survey of Polarization Asymmetries Predicted by {QCD}}",
    reportNumber = "IC-80-148",
    doi = "10.1002/prop.19810291102",
    journal = "Fortsch. Phys.",
    volume = "30",
    pages = "505--550",
    year = "1981"
}

@article{Artru:1989zv,
    author = "Artru, Xavier and Mekhfi, Mustapha",
    title = "{Transversely Polarized Parton Densities, their Evolution and their Measurement}",
    reportNumber = "LPTHE-ORSAY-89-25",
    doi = "10.1007/BF01556280",
    journal = "Z. Phys. C",
    volume = "45",
    pages = "669",
    year = "1990"
}

@article{Jaffe:1991kp,
    author = "Jaffe, R. L. and Ji, Xiang-Dong",
    title = "{Chiral odd parton distributions and polarized Drell-Yan}",
    reportNumber = "MIT-CTP-1952",
    doi = "10.1103/PhysRevLett.67.552",
    journal = "Phys. Rev. Lett.",
    volume = "67",
    pages = "552--555",
    year = "1991"
}

@article{Jaffe:1991ra,
    author = "Jaffe, R. L. and Ji, Xiang-Dong",
    title = "{Chiral odd parton distributions and Drell-Yan processes}",
    reportNumber = "MIT-CTP-2005",
    doi = "10.1016/0550-3213(92)90110-W",
    journal = "Nucl. Phys. B",
    volume = "375",
    pages = "527--560",
    year = "1992"
}

@inproceedings{Jaffe:1997yz,
    author = "Jaffe, R. L.",
    title = "{Can transversity be measured?}",
    booktitle = "{2nd Topical Workshop on Deep Inelastic Scattering off Polarized Targets: Theory Meets Experiment (SPIN 97)}",
    eprint = "hep-ph/9710465",
    archivePrefix = "arXiv",
    reportNumber = "MIT-CTP-2685",
    pages = "167--180",
    month = "10",
    year = "1997"
}

@article{Jaffe:1993xb,
    author = "Jaffe, R. L. and Ji, Xiang-Dong",
    title = "{Novel quark fragmentation functions and the nucleon's transversity distribution}",
    eprint = "hep-ph/9307329",
    archivePrefix = "arXiv",
    reportNumber = "MIT-CTP-2158",
    doi = "10.1103/PhysRevLett.71.2547",
    journal = "Phys. Rev. Lett.",
    volume = "71",
    pages = "2547--2550",
    year = "1993"
}

@article{Boer:1996ay,
    author = "Boer, Daniel and Tangerman, R. D.",
    title = "{Higher twist quark mass contributions to deep inelastic scattering}",
    eprint = "hep-ph/9603428",
    archivePrefix = "arXiv",
    reportNumber = "NIKHEF-96-016",
    doi = "10.1016/0370-2693(96)00541-2",
    journal = "Phys. Lett. B",
    volume = "381",
    pages = "305--310",
    year = "1996"
}

@article{Scopetta:1997qg,
    author = "Scopetta, Sergio and Vento, Vicente",
    title = "{A Quark model analysis of the transversity distribution}",
    eprint = "hep-ph/9706413",
    archivePrefix = "arXiv",
    reportNumber = "FTUV-97-29, IFIC-97-29",
    doi = "10.1016/S0370-2693(98)00212-3",
    journal = "Phys. Lett. B",
    volume = "424",
    pages = "25--32",
    year = "1998"
}

@article{Ruiz:2023ozv,
    author = "Ruiz, R. and others",
    title = "{Target mass corrections in lepton{\textendash}nucleus DIS: Theory and applications to nuclear PDFs}",
    eprint = "2301.07715",
    archivePrefix = "arXiv",
    primaryClass = "hep-ph",
    reportNumber = "IFJPAN-IV-2022-18, SMU-HEP-22-12, MS-TP-22-49, ANL-180568, FERMILAB-PUB-23-022-ND, JLAB-THY-23-3748",
    doi = "10.1016/j.ppnp.2023.104096",
    journal = "Prog. Part. Nucl. Phys.",
    volume = "136",
    pages = "104096",
    year = "2024"
}

@article{Barone:2001sp,
    author = "Barone, Vincenzo and Drago, Alessandro and Ratcliffe, Philip G.",
    title = "{Transverse polarisation of quarks in hadrons}",
    eprint = "hep-ph/0104283",
    archivePrefix = "arXiv",
    doi = "10.1016/S0370-1573(01)00051-5",
    journal = "Phys. Rept.",
    volume = "359",
    pages = "1--168",
    year = "2002"
}

@article{ParticleDataGroup:2026mpi,
    author = "Takahashi, F. and others",
    collaboration = "Particle Data Group",
    title = "{Review of Particle Physics}",
    doi = "10.1142/s0217751x26300115",
    journal = "Int. J. Mod. Phys. A",
    volume = "41",
    number = "22",
    pages = "2630011",
    year = "2026"
}

@article{Mulders:1995dh,
    author = "Mulders, P. J. and Tangerman, R. D.",
    title = "{The Complete tree level result up to order 1/Q for polarized deep inelastic leptoproduction}",
    eprint = "hep-ph/9510301",
    archivePrefix = "arXiv",
    reportNumber = "NIKHEF-95-053",
    doi = "10.1016/0550-3213(95)00632-X",
    journal = "Nucl. Phys. B",
    volume = "461",
    pages = "197--237",
    year = "1996",
    note = "[Erratum: Nucl.Phys.B 484, 538--540 (1997)]"
}

@article{Bacchetta:2006tn,
    author = "Bacchetta, Alessandro and Diehl, Markus and Goeke, Klaus and Metz, Andreas and Mulders, Piet J. and Schlegel, Marc",
    title = "{Semi-inclusive deep inelastic scattering at small transverse momentum}",
    eprint = "hep-ph/0611265",
    archivePrefix = "arXiv",
    reportNumber = "DESY-06-204",
    doi = "10.1088/1126-6708/2007/02/093",
    journal = "JHEP",
    volume = "02",
    pages = "093",
    year = "2007"
}

@inproceedings{Melnitchouk:2026isw,
    author = "Melnitchouk, W.",
    title = "{Higher twists in QCD}",
    eprint = "2609.19620",
    archivePrefix = "arXiv",
    primaryClass = "hep-ph",
    reportNumber = "JLAB-THY-26-4888",
    month = "9",
    year = "2026"
}

@article{Bacchetta:2008af,
    author = "Bacchetta, Alessandro and Conti, Francesco and Radici, Marco",
    title = "{Transverse-momentum distributions in a diquark spectator model}",
    eprint = "0807.0323",
    archivePrefix = "arXiv",
    primaryClass = "hep-ph",
    reportNumber = "JLAB-THY-08-841",
    doi = "10.1103/PhysRevD.78.074010",
    journal = "Phys. Rev. D",
    volume = "78",
    pages = "074010",
    year = "2008"
}

@article{Yang:2026cts,
    author = "Yang, Weihua",
    title = "{Twist-3 effects in fully inclusive jet production from $e^+e^-$ annihilation}",
    eprint = "2607.13525",
    archivePrefix = "arXiv",
    primaryClass = "hep-ph",
    doi = "10.1140/epjc/s10052-026-16232-8",
    journal = "Eur. Phys. J. C",
    volume = "86",
    number = "8",
    pages = "978",
    year = "2026"
}

@article{Accardi:2017pmi,
    author = "Accardi, Alberto and Bacchetta, Alessandro",
    title = "{Accessing the nucleon transverse structure in inclusive deep inelastic scattering}",
    eprint = "1706.02000",
    archivePrefix = "arXiv",
    primaryClass = "hep-ph",
    reportNumber = "JLAB-THY-17-2483",
    doi = "10.1016/j.physletb.2017.08.074",
    journal = "Phys. Lett. B",
    volume = "773",
    pages = "632--638",
    year = "2017"
}

@article{Accardi:2018gmh,
    author = "Accardi, Alberto and Signori, Andrea",
    title = "{Transversity in inclusive DIS and novel TMD sum rules}",
    eprint = "1808.00565",
    archivePrefix = "arXiv",
    primaryClass = "hep-ph",
    reportNumber = "JLAB-THY-18-2774",
    doi = "10.22323/1.316.0158",
    journal = "PoS",
    volume = "DIS2018",
    pages = "158",
    year = "2018"
}

@article{Mao:2014aoa,
    author = "Mao, Wenjuan and Lu, Zhun and Ma, Bo-Qiang",
    title = "{Transverse single-spin asymmetries of pion production in semi-inclusive DIS at subleading twist}",
    eprint = "1405.3876",
    archivePrefix = "arXiv",
    primaryClass = "hep-ph",
    doi = "10.1103/PhysRevD.90.014048",
    journal = "Phys. Rev. D",
    volume = "90",
    number = "1",
    pages = "014048",
    year = "2014"
}

@article{Mao:2014fma,
    author = "Mao, Wenjuan and Lu, Zhun and Ma, Bo-Qiang and Schmidt, Ivan",
    title = "{Double spin asymmetries $A_{LT}^{\cos\phi_S}$ and $A_{LT}^{\cos(2\phi_h -\phi_S)}$ in semi-inclusive DIS}",
    eprint = "1412.7390",
    archivePrefix = "arXiv",
    primaryClass = "hep-ph",
    doi = "10.1103/PhysRevD.91.034029",
    journal = "Phys. Rev. D",
    volume = "91",
    number = "3",
    pages = "034029",
    year = "2015"
}

@article{Accardi:2020iqn,
    author = "Accardi, Alberto and Signori, Andrea",
    title = "{On the connection between quark propagation and hadronization}",
    eprint = "2005.11310",
    archivePrefix = "arXiv",
    primaryClass = "hep-ph",
    reportNumber = "JLAB-THY-20-3193",
    doi = "10.1140/epjc/s10052-020-8380-1",
    journal = "Eur. Phys. J. C",
    volume = "80",
    number = "9",
    pages = "825",
    year = "2020"
}

@article{Afanasev:2007ii,
    author = "Afanasev, A. and Strikman, M. and Weiss, C.",
    title = "{Transverse target spin asymmetry in inclusive DIS with two-photon exchange}",
    eprint = "0709.0901",
    archivePrefix = "arXiv",
    primaryClass = "hep-ph",
    reportNumber = "JLAB-THY-07-717",
    doi = "10.1103/PhysRevD.77.014028",
    journal = "Phys. Rev. D",
    volume = "77",
    pages = "014028",
    year = "2008"
}

\end{document}